\documentclass[article]{elsarticle}

\usepackage{lineno,hyperref}

\usepackage{graphicx, epsfig}
\usepackage{float}
\usepackage{csquotes}
\usepackage{extarrows}
\usepackage{epstopdf}
\usepackage{bm}
\usepackage{bbm}
\usepackage{mathrsfs}
\usepackage{natbib}
\usepackage{caption}
\usepackage{subcaption}
\usepackage{xcolor}
\usepackage{multirow}
\definecolor{maroon}{cmyk}{0.74, 0.14, 0, 0.03}
\definecolor{mycolor}{HTML}{993333}
\newcommand{\ct}[1]{{\color{mycolor}\citet{#1}}}
\newcommand{\calt}[1]{{\color{mycolor}\citealt{#1}}}
\usepackage{amsfonts}
\usepackage[ruled,lined]{algorithm2e}
\newcommand{\tabincell}[2]{\begin{tabular}{@{}#1@{}}#2\end{tabular}}
\newcommand{\red}{\textcolor{red}}

\newtheorem{theorem}{Theorem}
 
\newtheorem{lemma}{Lemma}
\newtheorem{proposition}{Proposition}
\newtheorem{definition}{Definition}

\newtheorem{corollary}{Corollary}
\usepackage{geometry}
\usepackage{setspace}
\usepackage{pifont}
\usepackage{cancel}
 \bibpunct[, ]{(}{)}{,}{a}{}{,}%
\usepackage{amssymb}
\def\l{\left}\def\r{\right}
\def\prox{{\rm prox}}
\def\dom{{\rm dom}}
\def\x{\mathbf{x}}
\def\y{\mathbf{y}}
\def\e{\mathbf{e}}

\def\R{{\mathbb{R}}}
\def\S{{\mathbb{S}}}

\def\I{{\mathbb{I}}}
\def\bL{{\mathbb{L}}}

\def\cM {{\mathcal{M}}}
\def\cN {{\mathcal{N}}}
\def\cQ {{\mathcal{Q}}}
\def\cU {{\mathcal{U}}}

\def\cC {{\mathcal{C}}}
\def\cD {{\mathcal{D}}}

\def\cG {{\mathcal{G}}}

\newcommand{\norm}[1]{\left\lVert#1\right\rVert}
\newcommand{\blue}{\textcolor{black}}

\let\linenumbers\nolinenumbers\nolinenumbers

\graphicspath{{figures/}}
\begin{document}

\begin{frontmatter}


\title{Robust and Sparse Portfolio Selection: Quantitative Insights and Efficient Algorithms}


\author[mymainaddress]{Jingnan Chen}
\ead{jchen@buaa.edu.cn}

\author[mysecondaryaddress]{Selin~Damla~Ahipa\c{s}ao\u{g}lu}
\ead{ahipasa@gmail.com}

\author[mythirdaddress]{Ning Zhang}
\ead{zhangning@dgut.edu.cn}

\author[myfourthaddress]{Yufei Yang}
\ead{eeyufei@gmail.com}

\address[mymainaddress]{School of Economics and Management, Beihang University, Beijing, China}

\address[mysecondaryaddress]{School of Mathematical Sciences, University of Southampton, UK}

\address[mythirdaddress]{Corresponding author. School of Computer Science, Dongguan University of Technology, Dongguan, China}

\address[myfourthaddress]{Engineering Systems and Design, Singapore University of Technology and Design, Singapore}

\begin{abstract}
We extend the classical mean-variance (MV) framework and propose a robust and sparse portfolio selection model incorporating an ellipsoidal uncertainty set to reduce the impact of estimation errors and fixed transaction costs to penalize over-diversification. In the literature, the MV model under fixed transaction costs is referred to as the \emph{sparse} or \emph{cardinality-constrained} MV optimization, which is a mixed integer problem and is challenging to solve when the number of assets is large. We develop an efficient \emph{semismooth Newton-based proximal difference-of-convex algorithm} to solve the proposed model and prove its convergence to at least a local minimizer with a locally linear convergence rate. We explore properties of the robust and sparse portfolio  both analytically and numerically. In particular, we show that the MV optimization is indeed a robust procedure as long as an investor makes the proper choice on the risk-aversion coefficient. We contribute to the literature by proving that there is a one-to-one correspondence between the risk-aversion coefficient and the level of robustness. Moreover, we characterize how the number of traded assets changes with respect to the interaction between the level of uncertainty on model parameters and the magnitude of transaction cost.

\end{abstract}


\end{frontmatter}

\linenumbers

\section{Introduction}

The mean-variance (MV) framework, built by Markowitz to guide portfolio selection while considering both expected return and risk, is now considered an industrial benchmark. Modern Portfolio Theory based on the principles of Markowitz's framework inherently promotes diversification. 
Portfolio diversification is needed to alleviate the risks and stabilize portfolio weights. However, often this leads to over-diversification, where stocks are included into a portfolio merely to reduce the variance, sometimes sacrificing portfolio return.

We observe that some well-known investors prefer to work with concentrated portfolios.
This neglect of diversification, widely observed in practice, is known as the \lq\lq diversification paradox\rq\rq (\calt{Chha05}). Working with a concentrated portfolio can indeed facilitate better management while lowering costs associated with monitoring and \textcolor{black}{trading} assets.  For example, \ct{IvkovicSialmWeisbenner08} show that stock investments made by households that choose to concentrate their brokerage accounts in a few stocks outperform those made by households with more diversified accounts (especially among those with large portfolios).  The following quote by Warren Buffett resonates with expert practitioners:

\begin{displayquote}
	\lq\lq If you are a professional and have confidence, then I would advocate lots of concentration. \dots It's crazy to put money in your twentieth choice rather than your first.\rq\rq
\end{displayquote}

As the discussion above suggests, it is important to find a balance between diversification and concentration.
In this paper, we build a stylistic model, a generalization of the MV optimization given as follows:
\begin{equation}\label{rsmv0}
\text{RSMV}:=\left\{\begin{array}{cl}
\displaystyle\min_{\mathbf{x}\in\cC}\max_{{\mathbf{r}}} & \displaystyle\kappa\,\mathbf{x}^T\bm{\Sigma}\mathbf{x}-{\mathbf{r}}^T\mathbf{x}+\bm{\phi}^T\mathbbm{1}(\mathbf{x})\\\nonumber
\text{subject to} &\displaystyle ({\mathbf{r}}-\bar{\mathbf{r}})^T\bm{\Omega}^{-1}_{\bar{\mathbf{r}}}({\mathbf{r}}-\bar{\mathbf{r}})\leq \varepsilon,
\end{array}\right.
\end{equation}
where \blue{$\mathbf{x}$ is a vector with the $i$-th element representing the weight (i.e., the fraction of the total wealth held) of the $i$-th asset in the portfolio,  $\mathbf{r}$ is a vector of worse-case returns of the assets, $\kappa \geq 0$ is the risk-aversion coefficient, $\varepsilon \geq 0$} is the uncertainty level, $\bar{\mathbf{r}}$ and $\mathbf{\Sigma}$ denote the estimated \blue{model} parameters \blue{(i.e., estimated mean vector and covariance matrix of asset returns)}, $\bm{\Omega}_{\bar{\mathbf{r}}}$ is the estimation error covariance matrix of $\bar{\mathbf{r}}$, $\bm{\phi}$ 
 is a vector with the $i$-th element representing the $i$-th asset's fixed transaction cost, $\mathbf{e}$ is an all-one vector, $\mathbbm{1}(\mathbf{x})$ is a vector indicator function whose $i$-th element is equal to one if $x_i\neq 0$ and equal to zero otherwise, and the constraint set $\cC:=\{\x\in\R^n: \e^T\x-1=0\}$.

The \emph{Robust Sparse Mean Variance} (RSMV) model extends the classical MV framework incorporating robustness and sparsity through the ellipsidal uncertainty set and fixed transaction costs, respectively. The MV portfolio is known to be unstable with respect to estimated expectations of asset returns, where a slight perturbation may lead to a dramatic change in  portfolio weights (\calt{BesGra1991, JagMa2003, ChoZie1993}). 
Robust portfolio selection, where model parameters are specified to lie in uncertainty sets instead of being assigned to point values, has become a popular approach to mitigate the impact of estimation errors. Two most common choices for the uncertainty sets have been hypercubes and ellipsoids (\calt{GolIye2003, TutKoe2004, GarRamTan07, gregory2011robust, BoyGarUppWan2012, KimKimFab2014}). In particular, the ellipsoidal uncertainty set over the expectation of asset returns has an interesting connection with the worst-case value-at-risk (\calt{GhaOksOus2003, NatPacSim2008, ZymKuhRus2013}) and is hence adopted in the RSMV model.

The RSMV model considers the \emph{fixed transaction cost}, which is levied on each traded asset regardless of its position change (\calt{PatSub1982}), to avoid holding small portions of assets. In the literature, the MV model under fixed transaction costs is referred to as the \emph{sparse} or \emph{cardinality-constrained} MV optimization, which is a mixed integer problem and numerical methods have been developed to obtain near-optimal or optimal sparse portfolios. For example, \ct{LobFazBoy2007} describe an iterative reweighted algorithm to seek near-optimal sparse portfolios; the branch $\&$ bound algorithm and its variants have been tailored to calculate exact solutions (\calt{Bie1995, Sha2008, BerShi2009, GaoLi2013, ZheSunLi2014, bertsimas2022scalable}).

To dissect the structure of the RSMV portfolio, we first focus on the robust effect assuming zero transaction cost (\(\bm{\phi}=0\)). The corresponding robust portfolio is denoted as RMV portfolio. \ct{GarRamTan07} show that the RMV portfolio can be replicated by a convex combination of two benchmark portfolios:  the MV portfolio and the minimum-variance portfolio. We further demonstrate that the following three techniques have the same effect: i) choosing a larger risk-aversion coefficient $\kappa$, ii) using an ellipsoidal uncertainty set, or iii) shrinking the expected return towards the vector $\mathbf{e}$.  That is, the MV optimization is indeed a robust procedure as long as an investor makes the right choice on the risk-aversion coefficient $\kappa$. We contribute to the literature by proving that there is a one-to-one correspondence between the risk-aversion coefficient $\kappa$ and the uncertainty level $\varepsilon$, which could be used as a guideline on the choice of $\kappa$.

With transaction costs incorporated, we try to study and somehow `demystify' the ``diversification paradox" using both analytical and computational approaches. In particular, we want to understand whether it is always true that more assets must be included to  hedge against estimation errors and hence to improve the robustness of portfolios.  We find that this common perception is not necessarily always correct as in certain situations decreasing the number of assets can actually promote the robustness. Specifically, under a parameterized covariance matrix, we characterize conditions under which the cardinality of the portfolio weights might increase, decrease, or remain the same when $\varepsilon$ increases.  Although  diversity (i.e., including more assets) is needed to improve the portfolio stability in most cases, sometimes robust portfolios may be obtained by excluding some assets. 

Since the RSMV model is known to be NP-hard, we develop an efficient solution framework named \emph{Semismooth Newton-based proximal Difference-of-Convex Algorithm} (SN-pDCA) that obtains high-quality solutions for large-scale instances. Specifically, we propose a difference-of-convex (dc) problem to approximate the RSMV model. Then we introduce a proximal dc algorithm for the approximation problem, where the subproblems are solved using second-order information. With proper choice of parameters, we ensure that the global or local solution to the proposed approximation problem corresponds to the global or local minimizer of RSMV, respectively. We provide theoretical analyses to guarantee global convergence with a local linear convergence rate to the local minimizer of RSMV. It should be noted that the SN-pDCA is applicable to cardinality-constrained quadratic programs in general. These type of problems have wide applications in practice such as compressed sensing and gene selection in bioinformatics.

We evaluate the quality of the solution returned by the SN-pDCA with respect to the exact solution provided by CPLEX. In the comparison, we also include another commonly adopted benchmark portfolio, which is the solution to the convex optimization problem \eqref{L1MV} replacing the discontiuous term in the RSMV model \eqref{rsmv0} by the continuous weighted $\ell_1$-norm. Using datasets available in the {\it Fama-French Data Library\footnote{http://mba.tuck.dartmouth.edu/pages/faculty/ken.french/data\_library.html},} the relative error of the SN-pDCA solution is less than 10\(\%\) compared to the CPLEX solution, which is also less than half of the relative error of the benchmark \(\ell_1\) portfolio. With regard to the computational scalability, as the number of assets becomes 100, the computational time of CPLEX increases significantly and reaches 10 minutes.  In contrast, it takes less than 1 second for the SN-pDCA to generate a suboptimal solution. Thus, the SN-pDCA can provide a high-quality solution within an acceptable computational time.
Moreover, the set of nonzero positions in the SN-pDCA portfolio is a subset of that of the benchmark \(\ell_1\) portfolio. Hence, the computational speed of the SN-pDCA can be further improved if we reduce the model dimension according to the actively traded assets in the benchmark \(\ell_1\) portfolio.

The main contributions of this study are three-fold. First, we propose the RSMV model that extends the classical MV model by incorporating robustness and sparsity. We obtain a series of analytical results, providing qualitative guidance on portfolio investment. Second, we develop an efficient algorithm for solving large-scale RSMV model and demonstrate its convergence. Third, we evaluate the performance of the proposed SN-pDCA algorithm and verify the analytical properties via numerical examples. The remaining of the paper proceeds as follows. Section \ref{sec:Model} introduces the RSMV model and derives properties of the RSMV portfolio. Section 3 presents the solution framework and Section 4 reports numerical results. The paper is concluded in Section \ref{sec:Conc}. To ease exposition of our results, proofs are provided in the Appendix.

\textbf{Notation}: We use lowercase boldface letters to denote column vectors and uppercase boldface letters to denote matrices, e.g., $\mathbf{x}$ and $\mathbf{X}$. The space of symmetric matrices of dimension $n$ is denoted by $\mathbb{S}^n$. For any two matrices $\mathbf{X}, \mathbf{Y}\in\mathbb{S}^n$, we let $\langle\mathbf{X},\mathbf{Y}\rangle=\text{Tr}(\mathbf{X}\mathbf{Y})$ be the trace scalar product, whereas the relation $\mathbf{X}\succeq \mathbf{Y}$ ($\mathbf{X}\succ\mathbf{Y}$) implies that $\mathbf{X}-\mathbf{Y}$ is positive semidefinite (positive definite). We also denote $\mathbf{0}$ as the zero vector or matrix based on the context, and $\mathbf{I}$ as the identity matrix. We denote the Euclidean ($l_2$) norm  for a vector $\mathbf{x}\in\mathbb{R}^n$ as  $\norm{\cdot}_2$, i.e., $\norm{\mathbf{x}}_2=\sqrt{\mathbf{x}^T\mathbf{x}}$.




\section{Robust and Sparse MV Portfolio Optimization}\label{sec:Model}


In the RSMV model, we assume that the expectation of asset returns is confined to an ellipsoidal uncertainty set while the covariance matrix is known. The reason is that the mean-variance portfolio is more sensitive to the estimation errors in the mean of asset returns than in the covariance matrix.
In theory, $\bm{\Omega}_{\bar{\mathbf{r}}}$ equals to $\bm{\Sigma}$ if asset returns in a given sample are independent and identically distributed (\calt{Fab2007}). In most robust MV portfolio selection literature, $\bm{\Omega}_{\bar{\mathbf{r}}}$ is assumed to be a scaled version of $\bm{\Sigma}$ (\calt{GolIye2003, CerStu2006, GarRamTan07}) or a diagonal matrix (\calt{BoyGarUppWan2012, KimKimFab2014}).
When $\bm{\Omega}_{\bar{\mathbf{r}}}=\bm{\Sigma}$, the RSMV model can be rewritten as
\begin{equation}\label{rsmv}
\displaystyle\min_{\mathbf{x}\in\cC} \  \kappa\,\mathbf{x}^T\bm{\Sigma}\mathbf{x}+\sqrt{\varepsilon}\,\sqrt{\mathbf{x}^T\bm{\Sigma}\mathbf{x}}-\bar{\mathbf{r}}^T\mathbf{x}+\bm{\phi}^T\mathbbm{1}(\mathbf{x}).
\end{equation}
We can immediately see that the first term  is the variance of the portfolio $\mathbf{x}$ \blue{multiplied by the risk-aversion coefficient}; the second and third terms together coincide with the worst-case value-at-risk (WVaR) in \ct{GhaOksOus2003}, which is the largest VaR attainable among distributions with identical first and second order moment information; the last term is the total fixed transaction cost of the portfolio $\mathbf{x}$. Therefore, the objective of RSMV model is to select a portfolio by balancing its variance, WVaR, and fixed transaction costs. By investigating RSMV we will be able to characterize the impact of the uncertainty level and fixed transaction costs on the cardinality of a portfolio.

\subsection{Robust effects}\label{sec:V2R}
In this section, we focus on the robust MV optimization when \(\phi=0\), given by 
\begin{equation}\label{rmv}
\text{RMV}:=\min_{\mathbf{x}\in\cC} \  \kappa\,\mathbf{x}^T\bm{\Sigma}\mathbf{x}+\sqrt{\varepsilon}\,\sqrt{\mathbf{x}^T\bm{\Sigma}\mathbf{x}}-\bar{\mathbf{r}}^T\mathbf{x}.
\end{equation}

When $\varepsilon=0$ and $\bar{\mathbf{r}}=\mathbf{0}$,
the optimal solution is $\mathbf{x}_{\text{MIN}}=\frac{\bm{\Sigma}^{-1}\mathbf{e}}{\mathbf{e}^T\bm{\Sigma}^{-1}\mathbf{e}}$, known as the \emph{minimum-variance portfolio}, with optimal value equal to $v_\text{MIN}=1/\mathbf{e}^T\bm{\Sigma}^{-1}\mathbf{e}$.

When $\varepsilon=0$ but $\bar{\mathbf{r}}\neq \mathbf{0}$, the optimal solution is $\mathbf{x}_{\text{MV}}=\frac{1}{2}\widehat{\bm{\Sigma}}\bar{\mathbf{r}}+\mathbf{x}_\text{MIN}$, known as the \emph{mean-variance portfolio}, with optimal value ${v}_\text{MV}=\frac{\left(2\kappa-\mathbf{e}^T\bm{\Sigma}^{-1}\bar{\mathbf{r}}\right)^2}{4\kappa\mathbf{e}^T\bm{\Sigma}^{-1}\mathbf{e}}-\frac{\bar{\mathbf{r}}^T\bm{\Sigma}^{-1}\bar{\mathbf{r}}}{4\kappa}$, where $\widehat{\bm{\Sigma}}=\frac{1}{\kappa}\left(\bm{\Sigma}^{-1}-\frac{\bm{\Sigma}^{-1}\mathbf{e}\mathbf{e}^T\bm{\Sigma}^{-1}}{\mathbf{e}^T\bm{\Sigma}^{-1}\mathbf{e}}\right)$ is a positive semidefinite matrix.

Combining two different portfolio strategies is a popular approach to improve the out-of-sample performance (\calt{TuZhou2011, GarPed2013}). Our first proposition demonstrates that the RMV portfolio is equivalent to a convex combination of two benchmark portfolios, i.e.,  the \emph{mean-variance portfolio} and the \emph{minimum-variance portfolio}.
\begin{proposition}\label{thrm:VVR}
The $\text{RMV}$ portfolio in \eqref{rmv} is given by
\begin{eqnarray}
\displaystyle\mathbf{x}_{\emph{RMV}}&=&\frac{\kappa\rho^\star(\varepsilon)}{1+\kappa\rho^{\star}(\varepsilon)}\mathbf{x}_{\emph{MV}}+\frac{1}{1+\kappa\rho^{\star}(\varepsilon)}\mathbf{x}_{\emph{MIN}},\label{struct2}
\end{eqnarray}
where $\rho^{\star}(\varepsilon)>0$ is a 
monotone decreasing function of $\varepsilon$.
\end{proposition}

As $\rho^\star(\varepsilon)$ in equation (\ref{struct2}) is a monotone decreasing function of $\varepsilon$, we have $\mathbf{x}_{\text{RMV}}\rightarrow\mathbf{x}_{\text{MV}}$ as $\varepsilon\rightarrow 0$ and $\mathbf{x}_{\text{RMV}}\rightarrow\mathbf{x}_{\text{MIN}}$ as $\varepsilon\rightarrow \infty$, which indicates that an investor would rather use the minimum-variance strategy when there exists a high parameter uncertainty. Instead of solving the RMV problem \eqref{rmv} repeatedly, our result enables an investor to construct the $\mathbf{x}_{\text{RMV}}$ portfolio simply by taking a weighted average of the two benchmarks, where the weight reflects the investor's belief on the accuracy of the \blue{model parameter estimation}. The value of $\rho^\star(\varepsilon)$ can be easily evaluated by solving a quartic equation, or approximated by a closed-form formula (provided in Appendix).

Although this result is not new (see \calt{GarRamTan07}), our proof is not the same as it constructs the dual problem explicitly. This provides new insights to a well-known result. \blue{When $\kappa$ increases, the  $\mathbf{x}_{\text{RMV}}$ portfolio approaches the minimum-variance portfolio.}
When $\kappa=0$, the $\mathbf{x}_{\text{RMV}}$ calculates a WVaR portfolio.
\begin{proposition}\label{wwvar} For $\kappa=0$ and $\varepsilon>\bar{\mathbf{r}}^T\bm{\Sigma}^{-1}\bar{\mathbf{r}}-\frac{\left(\bar{\mathbf{r}}^T\bm{\Sigma}^{-1}{\mathbf{e}}\right)^2}{\mathbf{e}^T\bm{\Sigma}^{-1}\mathbf{e}}$, the RMV model \eqref{rmv} becomes
\begin{equation}\label{wvar}
\displaystyle\min_{\mathbf{x}\in \cC}\ \sqrt{\varepsilon}\sqrt{\mathbf{x}^T\bm{\Sigma}\mathbf{x}} - \bar{\mathbf{r}}^T\mathbf{x},
\end{equation}
and its optimal solution, denoted as the WVaR portfolio, is given by
$\mathbf{x}_{\emph{WVaR}}=\mathbf{x}_{\emph{MIN}}+\frac{\left(\bm{\Sigma}^{-1}-\frac{\bm{\Sigma}^{-1}\mathbf{e}\mathbf{e}^{T}\bm{\Sigma}^{-1}}{\mathbf{e}^T\bm{\Sigma}^{-1}\mathbf{e}}\right)\bar{\mathbf{r}}}{\sqrt{\left(\bar{\mathbf{r}}^T\bm{\Sigma}^{-1}\mathbf{e}\right)^2-\mathbf{e}^T\bm{\Sigma}^{-1}\mathbf{e}\left(\bar{\mathbf{r}}^T\bm{\Sigma}^{-1}\bar{\mathbf{r}}-\varepsilon\right)}}$,
with the optimal value being $
\frac{-\bar{\mathbf{r}}^T\bm{\Sigma}^{-1}\mathbf{e}+\sqrt{\left(\bar{\mathbf{r}}^T\bm{\Sigma}^{-1}\mathbf{e}\right)^2-\mathbf{e}^T\bm{\Sigma}^{-1}\mathbf{e}\left(\bar{\mathbf{r}}^T\bm{\Sigma}^{-1}\bar{\mathbf{r}}-\varepsilon\right)}}{\mathbf{e}^T\bm{\Sigma}^{-1}\mathbf{e}}$. 
\end{proposition}
Proposition \ref{wwvar} derives a closed-form formula for the WVaR portfolio that is equivalent to the MV portfolio with $\kappa$ being $\frac{\sqrt{\left(\bar{\mathbf{r}}^T\bm{\Sigma}^{-1}\mathbf{e}\right)^2-\mathbf{e}^T\bm{\Sigma}^{-1}\mathbf{e}\left(\bar{\mathbf{r}}^T\bm{\Sigma}^{-1}\bar{\mathbf{r}}-\varepsilon\right)}}{2}$.

In the following, we discuss some interesting connections of the RMV model with shrinkage estimators and other widely adopted portfolio models.
\begin{itemize}
\item \emph{The equivalence between risk-aversion coefficient and shrinkage estimator}: the RMV portfolio could be further expressed as $\frac{\kappa\rho^\star(\varepsilon)}{2(1+\kappa\rho^\star(\varepsilon))}\widehat{\bm{\Sigma}}\bar{\mathbf{r}}+\mathbf{x}_{\text{MIN}}$. Comparing it with the MV portfolio $\frac{1}{2}\widehat{\bm{\Sigma}}\bar{\mathbf{r}}+\mathbf{x}_{\text{MIN}}$, we can readily observe that the RMV portfolio is actually a mean-variance portfolio with the risk-aversion coefficient $\tilde{\kappa}$ given by $\tilde{\kappa}=\kappa+\frac{1}{\rho^\star(\varepsilon)}$ and $\tilde{\kappa}\rightarrow\infty$ \textcolor{black}{when $\epsilon\rightarrow\infty$, }as $\rho^\star(\varepsilon)$ is a monotone decreasing function. We could also obtain the same RMV portfolio by simply plugging the shrinkage estimator on the mean of asset returns, which is in the form of  $\frac{\kappa\rho^\star(\varepsilon)}{1+\kappa\rho^\star(\varepsilon)}\bar{\mathbf{r}}+\frac{1}{1+\kappa\rho^\star(\varepsilon)}v\mathbf{e}$, into the MV model, where the ratio $\frac{\kappa\rho^\star(\varepsilon)}{1+\kappa\rho^\star(\varepsilon)}$ is referred to as \emph{shrinkage intensity} (\calt{Jor86}) and $v$ is a scaling factor.

Therefore, we demonstrate that the following three techniques have the same effect: i) choosing a larger risk-aversion coefficient $\kappa$, ii) using an ellipsoidal uncertainty set, or iii) shrinking the expected return towards the target expected return $v\mathbf{e}$. Our results suggest that the MV optimization is indeed a robust procedure as long as an investor makes the right choice on the risk-aversion coefficient $\kappa$. We contribute to the literature by proving that there is a one-to-one correspondence between the risk-aversion coefficient $\kappa$ and the uncertainty level $\varepsilon$, which could be used as a guideline on the choice of $\kappa$.

 \item \emph{A unified framework}:
We have shown that RMV portfolio generalizes a set of well-studied portfolios. Each of these can be obtained as a combination of two benchmark portfolios as shown in Table \ref{tab1}.
\end{itemize}

\begin{table}[t]
\centering
\begin{tabular}{ll}
\hline
Portfolio Strategy & $\alpha_\text{Port}$\\ \hline
Mean-Variance &$\alpha_\text{MV}=1$ \\
Minimum-Variance &$\alpha_\text{MIN}=0$ \\
Worst-Case VaR &$\alpha_{\text{WVaR}}=\frac{2\kappa}{\sqrt{\left(\bar{\mathbf{r}}^T\bm{\Sigma}^{-1}\mathbf{e}\right)^2-\mathbf{e}^T\bm{\Sigma}^{-1}\mathbf{e}\left(\bar{\mathbf{r}}^T\bm{\Sigma}^{-1}\bar{\mathbf{r}}-\varepsilon\right)}}$ \\
Robust Mean-Variance &$\alpha_{\text{RMV}}=\frac{\kappa\rho}{1+\kappa\rho}, \rho \in [0, \infty)$\\\hline
\end{tabular}\vspace{0.5cm}
\caption{A unified formula for $\mathbf{x}_{\text{Port}}=\frac{\alpha_{\text{Port}}}{2}\widehat{\bm{\Sigma}}\bar{\mathbf{r}}+\frac{\bm{\Sigma}^{-1}\mathbf{e}}{\mathbf{e}^T\bm{\Sigma}^{-1}\mathbf{e}}$, where Port=$\{$MV, MIN, RMV, WVaR$\}$.}\label{tab1}
\end{table}

\begin{figure}[h]
 	\begin{center}
		\includegraphics[scale=0.6]{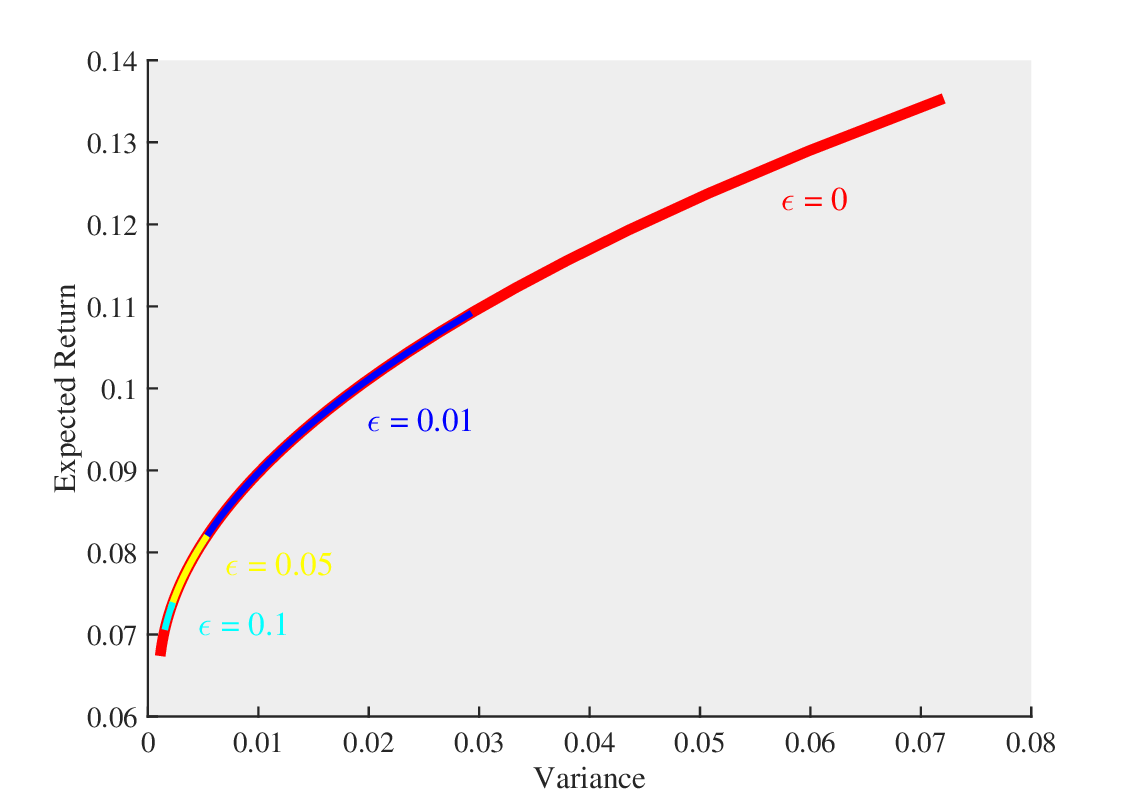}
		\caption {MV and RMV efficient frontiers}\label{fig:frontier}
 	\end{center}
\end{figure}

\noindent\textbf{Example 1:} Consider a market with three assets whose expected return vector is $[0.107,0.737,0.627]^T$ and covariance matrix is
	\begin{equation*}
	\left[ \begin{array}{ccc}
	0.02778&   0.00387&  0.00021\\
	0.00387&   0.01112 & -0.0002\\
	0.00021&  -0.0002  & 0.00115\\
	\end{array}\right].
	\end{equation*}
	In Figure \ref{fig:frontier}, we illustrate the mean-variance efficient frontier in the red thick curve with $\kappa$ ranging from $0.5$ to $10$. Specifically, we solve
	$\displaystyle\min_{\mathbf{x}:\,\mathbf{e}^T\mathbf{x}=1} \kappa\,\mathbf{x}^T\bm{\Sigma}\mathbf{x}-\bar{\mathbf{r}}^T\mathbf{x}
	$ for each $\kappa$ and then plot the pair ($\mathbf{x}^T_{\text{MV}}\bm{\Sigma}\mathbf{x}_\text{MV}$, $\bar{\mathbf{r}}^T\mathbf{x}_\text{MV}$). Similarly, we illustrate the RMV efficient frontier with $\kappa$ ranging from $0.5$ to $1.5$ and a given $\varepsilon$. For example, the blue, yellow, and cyan thin curves correspond to the RMV efficient frontier with $\varepsilon=0.01$, $0.05$, and $0.1$, respectively. We can observe that the three instances of the RMV efficient frontiers are simply parts of the MV efficient frontier, which verifies our discussions that the RMV portfolio is nothing but a MV portfolio with a larger $\kappa^\prime$ being $\kappa+\frac{1}{\rho^\star(\varepsilon)}$.

\subsection{Diversification paradox}\label{sec:SimBnd}
In this section, we aim to understand how the number of assets changes with respect to different uncertainty levels and transaction costs. For analytical tractability, in the sequel, we consider the case that the covariance matrix in the ellipsoidal uncertainty set is $\bm{\Omega}_{\bar{\mathbf{r}}}=\textbf{I}$, where \textbf{I} is the identity matrix. The RMV model \eqref{rmv} becomes the \(\ell_2\)-regularized MV model: \(\displaystyle\min_{\mathbf{x}\in \cC}\  \kappa\,\mathbf{x}^T\bm{\Sigma}\mathbf{x}-\bar{\mathbf{r}}^T\mathbf{x}+\sqrt{\varepsilon}\norm{\mathbf{x}}_2\). We further replace $\norm{\mathbf{x}}_2$ with $\norm{\mathbf{x}}_2^2$ as in \ct{DGNU09}, while most insights obtained can be applied to models under general covariance matrices as illustrated in the numerical examples. The modified RMV model is given by

\begin{equation}\label{mrmv}
\displaystyle \text{RMV}^{l_2}:= \min_{\mathbf{x}\in \cC} \  \kappa\,\mathbf{x}^T\bm{\Sigma}\mathbf{x}-\bar{\mathbf{r}}^T\mathbf{x}+\sqrt{\varepsilon}\norm{\mathbf{x}}_2^2=\kappa\,\mathbf{x}^T\left(\bm{\Sigma}+\frac{\sqrt{\varepsilon}}{\kappa}\mathbf{I}\right)\mathbf{x}-\bar{\mathbf{r}}^T\mathbf{x},
\end{equation}
whose optimal solution is $\mathbf{x}_{\text{MV}}^{l_2}=\frac{1}{2\kappa}\left(\widetilde{\bm{\Sigma}}^{-1}-\frac{\widetilde{\bm{\Sigma}}^{-1}\mathbf{e}\mathbf{e}^T\widetilde{\bm{\Sigma}}^{-1}}{\mathbf{e}^T\widetilde{\bm{\Sigma}}^{-1}\mathbf{e}}\right)\bar{\mathbf{r}}+\frac{\widetilde{\bm{\Sigma}}^{-1}}{\mathbf{e}^T\widetilde{\bm{\Sigma}}^{-1}\mathbf{e}}\mathbf{e}$, known as the $l_2$-regularized MV portfolio, with $\widetilde{\bm{\Sigma}}=\bm{\Sigma}+\left(\sqrt{\varepsilon}/{\kappa}\right)\mathbf{I}$. Similarly, the modified RSMV model becomes
\begin{equation}\label{mrsmv}
\displaystyle \text{RSMV}^{l_2}:= \min_{\mathbf{x}\in \cC}  \ \kappa\,\mathbf{x}^T\bm{\Sigma}\mathbf{x}-\bar{\mathbf{r}}^T\mathbf{x}+\sqrt{\varepsilon}\norm{\mathbf{x}}_2^2+\bm{\phi}^T\mathbbm{1}(\mathbf{x}).
\end{equation}

We first show that the $l_2$-regularized MV portfolio converges to the $1/N$ portfolio (denoted by $\mathbf{x}_\text{EW}$ below) at a rate of  $\mathcal{O}(1/\sqrt{\varepsilon})$ and establish its equivalence to the combination rule in \ct{TuZhou2011}. By considering a parameterized covariance matrix, we then conduct a sensitivity analysis on the cardinality of the MV portfolio under transaction costs.


\begin{proposition}\label{coro1}
	The Euclidean distance between $\mathbf{x}_\emph{MV}^{l_2}$ and $\mathbf{x}_\emph{EW}$ is upper bounded by
	\begin{equation*}
	\displaystyle\norm{\mathbf{x}_\emph{MV}^{l_2}-\mathbf{x}_\emph{EW}}_2\leq\frac{c}{\lambda_{[N]}+\sqrt{\varepsilon}/\kappa}
	\end{equation*}
	where $\bm{\Sigma}=\mathbf{U}\bm{\Lambda}\mathbf{U}^T$ is the eigenvalue decomposition of $\bm{\Sigma}$ such that $\mathbf{U}\mathbf{U}^T=\mathbf{I}$ and $\bm{\Lambda}=\emph{diag}(\bm{\lambda})$, $\lambda_{[1]}\geq \lambda_{[2]}\geq\cdots\lambda_{[N]}$, and $ c=\frac{\norm{\bar{\mathbf{r}}}_2}{2\kappa}+\frac{\lambda_{[1]}\left(\lambda_{[1]}-\lambda_{[N]}\right)}{\sqrt{N}\lambda_{[N]}}$.
\end{proposition}
The  \(\ell_2\) regularizer has been shown to have equalizing effect (\calt{DGNU09,chen2020application}), and hence the \(\ell_2\)-regularized MV portfolio tends to be relatively close to the $1/N$ portfolio. Here we further prove that the convergence rate of $\mathbf{x}^{l_2}_\text{MV}$  to the $1/N$ portfolio is $\mathcal{O}(1/\sqrt{\varepsilon})$. This suggests that an investor gradually shifts the optimal mean-variance strategy to the naive diversification when there is high uncertainty in the estimated parameters. The gain from the mean-variance diversification is mostly offset by the estimation error as the $1/N$ portfolio ignores the prior information on the expectation of asset returns.

\ct{TuZhou2011} consider a combined portfolio
\begin{equation*}
\mathbf{x}_\text{c}=\beta\mathbf{x}_\text{EW}+(1-\beta)\mathbf{x}_\text{MV},
\end{equation*}
where $0\leq\beta\leq 1$ is the combination coefficient and determined by optimizing some expected loss function. The combined portfolio is shown to have a significant impact in improving the MV strategy and outperforms the $1/N$ portfolio in most scenarios. In this case, the Euclidean distance between the combined portfolio $\mathbf{x}_\text{c}$ and the $1/N$ portfolio is given by $(1-\beta)\norm{\mathbf{x}_\text{MV}-\mathbf{x}_\text{EW}}_2$. \textcolor{black}{Therefore, choosing $\varepsilon$ by solving  the following equation
\begin{equation*}
\displaystyle(1-\beta)\norm{\mathbf{x}_\text{MV}-\mathbf{x}_\text{EW}}_2=\frac{c}{\lambda_{[N]}+\sqrt{\varepsilon}/\kappa},
\end{equation*}
our $l_2$-regularized portfolio can have a similar performance as the combined portfolio of \ct{TuZhou2011}.}

Proposition \ref{coro1} shows the impact of the uncertainty level on the portfolio composition. In the following, we further investigate the joint effect of the uncertainty level and the transaction cost on the portfolio cardinality through a case study.

\begin{corollary}\label{co1}
	When \(\bar{\mathbf{r}}=0\) and \(\epsilon=0\), the Euclidean distance between the minimum-variance portfolio $\mathbf{x}_\emph{MIN}$ and the equal-weighted portfolio $\mathbf{x}_\emph{EW}$ is upper bounded by
	\begin{equation*}
	\displaystyle\norm{\mathbf{x}_\emph{MIN}-\mathbf{x}_\emph{EW}}_2\leq\frac{1}{N}cond(\boldsymbol{\Sigma})[cond(\boldsymbol{\Sigma})-1],
	\end{equation*}
	where \(cond(\boldsymbol{\Sigma})\) is the conditional number (the ratio of the maximum eigenvalue to the smallest eigenvalue) of the covariance matrix \(\boldsymbol{\Sigma}\).
\end{corollary}
Corollary \ref{co1} implies that the minimum-variance portfolio is diversified when the covariance matrix \(\boldsymbol{\Sigma}\) is well-conditioned. However, for a large-size portfolio, it might be difficult to find enough observations for estimating sample covariance matrix, which possibly leads to an ill-conditioned covariance matrix. In this case it is more important to incorporate the robust uncertainty set to improve model stability.

\subsubsection{Cardinality surface: a case study}
{For ease of exposition}, we consider the following parameterized covariance matrix (\calt{BoyGarUppWan2012})
\begin{equation}\label{scov}
\bm{\Sigma}\left(\sigma,\rho\right)=\left[\begin{array}{cccc}
\sigma^2&\rho\sigma^2&\cdots&\rho\sigma^2\\
\rho\sigma^2&\sigma^2&\cdots&\rho\sigma^2\\
\vdots&\vdots&\ddots&\vdots\\
\rho\sigma^2&\rho\sigma^2&\cdots&\sigma^2
\end{array}\right] = \sigma^2(1-\rho)\left(\mathbf{I}+\frac{\rho}{1-\rho}\mathbf{e}\mathbf{e}^T\right),
\end{equation}
where $\bm{\Sigma}\left(\sigma,\rho\right)$ is an approximation of $\bm{\Sigma}$ with $\sigma$ and $\rho$ obtained by solving a simple nearest matrix problem
\begin{equation*}
\min_{\sigma, -1\leq\rho\leq 1}\norm{\bm{\Sigma}\left(\sigma,\rho\right)-\bm{\Sigma}}_F^2.
\end{equation*}
In addition, we assume that \(\bm{\phi}=\phi \mathbf{e}\), where \(\phi\) is a positive constant. Substituting $\bm{\Sigma}\left(\sigma,\rho\right)$ into the modified RSMV model (\ref{mrsmv}), we obtain a set optimization problem
\begin{equation}\label{cardprob}
\displaystyle\min_{\mathcal{S}\subset\{1,2,\cdots,N\}}\frac{\kappa\sigma^2(1-\rho+\rho|\mathcal{S}|+\delta)}{|\mathcal{S}|}+\frac{(\mathbf{e}^T\bar{\mathbf{r}}_{\mathcal{S}})^2-|\mathcal{S}|\norm{\bar{\mathbf{r}}_{\mathcal{S}}}_2^2}{4\kappa\sigma^2(1-\rho+\delta)|\mathcal{S}|}-\frac{\mathbf{e}^T\bar{\mathbf{r}}_\mathcal{S}}{|\mathcal{S}|}+\phi|\mathcal{S}|,
\end{equation}
where $\delta=\sqrt{\varepsilon}/(\kappa\sigma^2)$, $\mathcal{S}=\{i: x_i\neq 0\}$ is the index set for the traded assets, and $|\mathcal{S}|$ denotes the cardinality of $\mathcal{S}$. Note that $\mathbf{e}^T\bar{\mathbf{r}}_\mathcal{S}\leq \sum_{i=1}^{|\mathcal{S}|}\bar{r}_{[i]}$ and $\left(\mathbf{e}^T\bar{\mathbf{r}}_\mathbf{S}\right)^2\leq|\mathcal{S}|\norm{\bar{\mathbf{r}}_\mathcal{S}}_2^2$. It is easy to verify that (\ref{cardprob}) is upper bounded by the following univariate problem
\begin{equation}\label{cardprobub}
\displaystyle\min_{s: s\in\{1,2,\cdots,N\}}v_\text{U}(s;\delta,\phi)=\frac{\kappa\sigma^2(1-\rho+\rho s+\delta)}{s}-\frac{\sum_{i=1}^{s}\bar{r}_{[i]}}{s}+\phi s,
\end{equation}
where the investor adopts a $1/s$ diversification strategy that chooses the first $s$ assets with the highest expected return, i.e., $\bar{r}_{[1]}\geq\bar{r}_{[2]}\geq\cdots\geq\bar{r}_{[s]}$, and the problem is to decide the number of assets to be included in the portfolio. This upper bound facilitates the investor to predict the trend of $s$ when $\delta$ or $\phi$ increases. We next discuss how the number of traded assets ($s$) changes with the uncertainty level under a fixed $\phi$.

\begin{proposition}\label{conds}
	Assume that $s^\star$ and $s^\prime$ are the optimal solutions of (\ref{cardprobub}) under the parameters ($\delta$, $\phi$) and ($\delta+\Delta$, $\phi$), respectively. The condition that $s^\prime$ is smaller than $s^\star$ is given by
	\begin{equation*}\label{c1}
	\text{C1}: 0<\Delta\leq \min\{B_{-}(s^\star), B_{+}(s^\star)\},
	\end{equation*}
	the condition that $s^\prime$ is equal to $s^\star$ is given by
	\begin{equation*}\label{c2}
	\text{C2}: \max\{0, B_{-}(s^\star)\}\leq \Delta\leq B_{+}(s^\star),
	\end{equation*}
	and the condition that $s^\prime$ is greater than $s^\star$ is given by
	\begin{equation*}\label{c3}
	\text{C3}: \Delta\geq \max\{0, B_{-}(s^\star), B_{+}(s^\star)\},
	\end{equation*}
	where
	\begin{equation*}
	\begin{array}{cl}
	&\displaystyle B_{-}(s^\star)=\rho-\delta-1-\frac{1}{\kappa\sigma^2}\min_{l<s^\star}\left(\frac{l\sum_{i=1}^{s^\star}\bar{r}_{[i]}-s^\star\sum_{i=1}^{l}\bar{r}_{[i]}}{s^\star-l}-\phi s^\star l\right),\\
	\text{~and~}&\displaystyle B_{+}(s^\star)=\rho-\delta-1+\frac{1}{\kappa\sigma^2}\min_{l>s^\star}\left(\frac{l\sum_{i=1}^{s^\star}\bar{r}_{[i]}-s^\star\sum_{i=1}^{l}\bar{r}_{[i]}}{l-s^\star}+\phi s^\star l\right).
	\end{array}
	\end{equation*}
\end{proposition}
In Proposition \ref{conds}, we show that (i) a slight increase of the uncertainty level (C1) could decrease the number of traded assets. It could happen when the cost of adding one more asset is higher than the risk reduction and profit enhancement. (ii) The number of traded assets will remain the same when there is only a mild increase of the uncertainty level (C2). The investor is confident that her current portfolio strategy is robust against a mild estimation error on the \blue{model} parameters and thus she is reluctant to alter the current portfolio composition by introducing more assets to further reduce the risk. (iii) A significant increase of the uncertainty level (C3) could result in a further diversified strategy as we have shown that the investor would like to adopt the $1/N$ diversification strategy when there is a high degree of estimation errors on the \blue{model} parameter in Proposition \ref{coro1}.

Note that when \(\delta\) is large and \(\varepsilon\) is small, only C3 is valid. This implies that in the presence of high parameter uncertainty and low transaction costs, investors are more inclined to adopt the $1/N$ diversification strategy, which aligns with our intuition. Conversely, if $\varepsilon$ is large or $\delta$ is small, investors must strike a balance between diversification and the associated transaction costs. Consequently, the number of traded assets could, in some cases, increase, decrease, or remain unchanged. 

{Though Proposition \ref{conds} is derived under the parametrized covariance matrix given in (\ref{scov}), the insights obtained here apply to a general setting of covariance matrices; refer to the computational results in Section \ref{Cardinality surface} for details. }  

\begin{figure}[t]
 	\begin{center}
 		\includegraphics[scale=1]{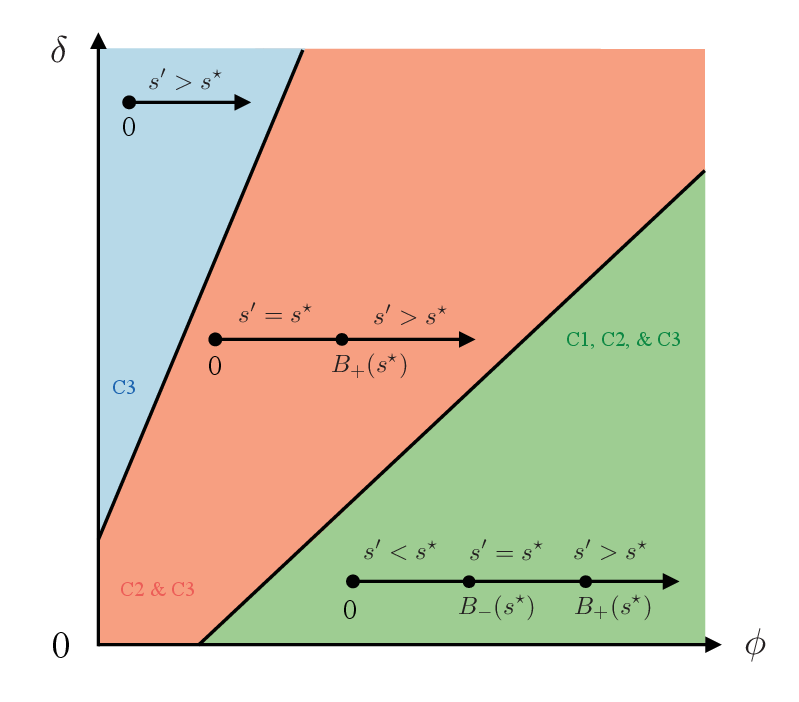}
 		\caption{The feasible regions for C1-C3}\label{fig:card}
 	\end{center}
\end{figure}

\noindent\textbf{Example 2:}
	Assume $r_{[i]}=\bar{r}-\Delta \bar{r}(i-1)$ where $0<\Delta \bar{r}<\bar{r}/(N-1)$, then $B_{\pm}(s^\star)=\rho-\delta-1+\frac{s^\star(s^\star\pm 1)\left({\Delta \bar{r}}/{2}+\phi\right)}{\kappa\sigma^2}$. The feasible triple ($\Delta$, $\delta$, $\phi$) for C1 - C3 is illustrated in Figure \ref{fig:card}.

\section{Solution Algorithm}
In this section, we develop an efficient algorithm to solve the large-scale RSMV model \eqref{rsmv} using a dc approximation approach.
We will demonstrate that the proposed approximation approach is capable of obtaining at least a local minimizer of the RSMV model \eqref{rsmv} with a locally linear convergence rate.

\subsection{Dc approximation}
To begin with, we rewrite the RSMV model \eqref{rsmv} as the following equivalent optimization problem:
\begin{equation}\label{L2CP}
\begin{array}{cl}
\min\limits_{\x\in\cC} & \frac{1}{2}\|\mathbf{Wx}\|^2_2+\lambda \|\mathbf{Wx}\|_2-\tilde{\mathbf{r}}^T\mathbf{x}+\tilde{\bm{\phi}}^T\mathbbm{1}(\mathbf{x}),
\end{array}
\end{equation}
where $\bm{\Sigma}=\mathbf{W}^T\mathbf{W}$ is the Cholesky decomposition of $\bm{\Sigma}$, $\lambda=\frac{\sqrt{\varepsilon}}{2\kappa}$, $\tilde{\mathbf{r}}=\frac{1}{2\kappa}\bar{\mathbf{r}}$, and $\tilde{\bm{\phi}}=\frac{1}{2\kappa}\bm{\phi}$. Note that model \eqref{L2CP} belongs to the class of cardinality-constrained quadratic programs, which has wide applications such as sparse signal representation in compressed sensing and gene selection in bioinformatics.

Due to the inherent discrete structure of $\mathbbm{1}(\x)$, model \eqref{L2CP} is classified as NP-hard. Over recent years, the convex $\ell_1$ penalty has often been utilized as a surrogate for $\mathbbm{1}(\x)$, and several iterative methods have been employed to address the convex $\ell_1$ penalized problem. However, the inclusion of the $\ell_1$ penalty frequently leads to a notable bias in the resulting estimator (\calt{fan2001variable}). To mitigate this issue, alternative nonconvex functions such as the smoothly clipped absolute deviation penalty (\calt{fan2001variable}), the minimax concave penalty function (\calt{zhang2010nearly}), and the capped-$\ell_1$ function (\calt{zhang2010analysis}) have been proposed to serve as surrogates for $\mathbbm{1}(\x)$. These nonconvex penalties have been shown to offer desirable traits including unbiasedness, data continuity, and sparsity properties.

\begin{figure}[h]
\centering
\includegraphics[scale=0.55]{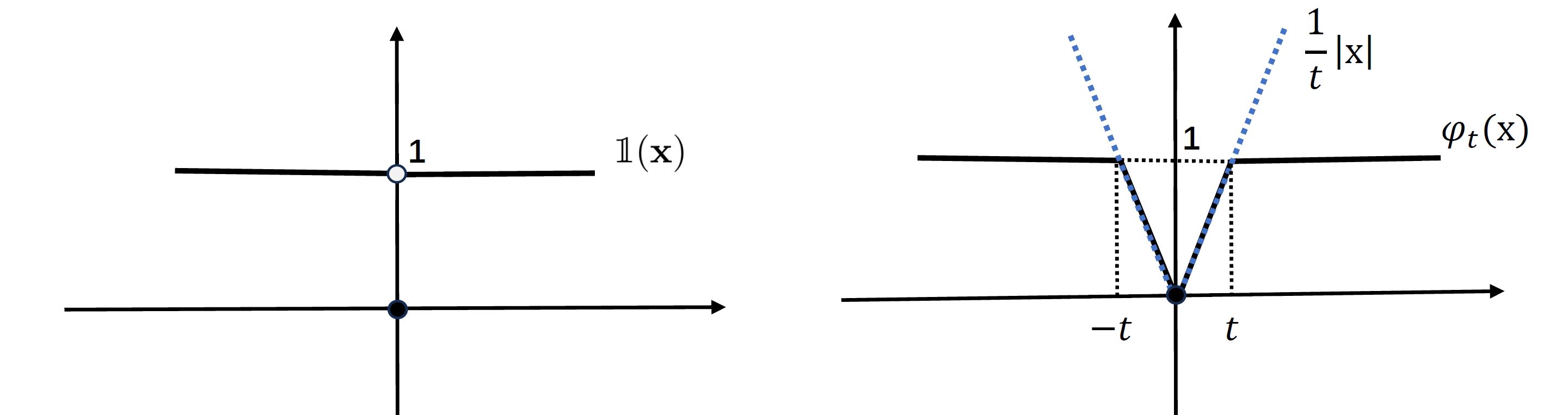}
\caption{One dimensional illustrations of $\mathbbm{1}(\x)$, the $\ell_1$ function $|\x|/t$, and the capped-$\ell_1$ function $\varphi_t(\x)$.}\label{fig-toy}
\end{figure}

{Among the set of candidate surrogates for \(\mathbbm{1}(\x)\) that can be expressed as the difference of two convex functions (\calt{ahn2017difference}), we select the continuous capped-\(\ell_1\) function introduced by \ct{zhang2010analysis}. This choice is motivated by its piecewise linear structure, which ensures the Kurdyka-{\L}ojasiewicz property with exponent \(1/2\) (see Definition \ref{def-KL}). This property is essential for analyzing the convergence rate of widely used first-order numerical methods. Recall that the capped-\(\ell_1\) function can be represented in a dc form:}
$$
\varphi_t(\x)=p_t(\x)-q_t(\x),
$$
where
$$
p_t(\x)=\frac{1}{t} \sum^n_{i=1}\tilde{\bm{\phi}}_i|\x_i|\quad\text{and}\quad q_t(\x)= \sum^n_{i=1}\tilde{\bm{\phi}}_i \max\{0,\x_i/t-1,-\x_i/t-1\}.
$$
Since both $p_t(\x)$ and $q_t(\x)$ are convex functions, $\varphi_t(\x)$ has a dc structure. Note that $p_t(\x)$ is a weighted $\ell_1$ norm and using $p_t(\x)$  to approximate $\tilde{\bm{\phi}}^T\mathbbm{1}(\mathbf{x})$ is a widely adopted approach. Figure \ref{fig-toy} provides a one-dimensional illustration of the capped-$\ell_1$ function $\varphi_t(\x)$. According to Figure \ref{fig-toy}, $\varphi_t(\x)$ provides a better approximation to the discrete function $\mathbbm{1}(\x)$ compared to the $\ell_1$ function $p_t(\x)$. The superior performance of the capped-$\ell_1$ approximation is also verified by the numerical examples in Section 4.
Consequently, we obtain the following continuous approximation of model \eqref{L2CP}:
\begin{equation}\label{uL2CP}
\begin{array}{cl}
\min\limits_{\x\in\cC}& \frac{1}{2}\|W\x\|^2_2+\lambda \|W\x\|_2-\tilde{\mathbf{r}}^T\mathbf{x}+ p_t(\x)-q_t(\x)
\end{array}
\end{equation}

For comparison, we also consider the following $\ell_1$ approximation model:
\begin{equation}\label{L1MV}
\begin{array}{cl}
\min\limits_{\x\in \cC}& \frac{1}{2}\|W\x\|^2+\lambda \|W\x\|-\tilde{\mathbf{r}}^T\mathbf{x}+ p_t(\x).
\end{array}
\end{equation}

{The \( \ell_1 \)-regularized model \eqref{L1MV}, abbreviated as the L1MV model, serves as a standard benchmark in nonconvex sparse optimization due to its convexity, computational efficiency, and effective sparsity induction, which together make it both practical and interpretable (see, e.g., \calt{tibshirani1996regression, brodie2009sparse, fastrich2015constructing, chen2022distributionally, zhang2022portfolio}). To evaluate the effectiveness of the dc approximation model \eqref{L2CP} and its solution approach, we adopt the L1MV model \eqref{L1MV} as a comparative benchmark.}



\subsection{Connections between the RSMV model and its dc approximation}

Since there is a lack of efficient numerical methods for finding the global solution of the RSMV model \eqref{L2CP} in high dimensions, we investigate the relationship between the local minimizer of the dc approximation \eqref{uL2CP} and that of model \eqref{L2CP}. The discussion in this section aligns with the framework presented in Section 2 of {\ct{bian2020smoothing}}. However, the inclusion of the equality constraint introduces additional challenges for theoretical analysis.
These challenges primarily stem from the characterization of the normal cone (as defined on Page 15 in \ct{rockafellar1996convex}) associated with the set of linear constraints, which no longer coincides with the set of zeros in \ct{bian2020smoothing}. Consequently, the optimality results in \ct{bian2020smoothing} cannot be directly applied in our context.

We first recall the definition of the lifted stationary point for model \eqref{uL2CP}, as originally proposed by \ct{pang2017computing} and adapted for the capped-$\ell_1$ regularized problem in {\color{mycolor}\cite{bian2020smoothing}} (see Definition 2.1). Let $\theta_1(s)=0,\,\theta_2(s):=s/t-1,\,\theta_3(s):=-s/t-1$. For $s\in\R$, we define the index set
\begin{equation*}
\cD(t) :=\l\{i\in\{1,2,3\}:\theta_i(s)=\max\{\theta_1(s),\theta_2(s),\theta_3(s)\}\r\}.
\end{equation*}
Moreover, consider
$$h(\x):=\frac{1}{2}\|Wx\|^2+\lambda\|Wx\|-\tilde{\mathbf{r}}^T\mathbf{x}$$
with a Lipschitz constant denoted by $L_h$. The normal cone associated with the set $\cC$ is given by
$\cN_{\cC}(\x)=\{s\e : s\in\R\}$ if $\x\in\cC$. Now we are ready to present the definition of the lifted stationary point.
\begin{definition}\label{def-lifted}
We say that $\x\in\cC$ is a lifted stationary point of \eqref{uL2CP} if there exist $d_i\in\cD(\x_i), i=1,\ldots,n$ such that
\begin{equation}\label{cond-lift}
\sum^n_{i=1}\tilde{\phi}_i\theta'_{d_i}(\x_i)\e_i\in \nabla h(\x)+\frac{1}{t}\sum^n_{i=1} \partial(\tilde{\phi}_i|\x_i|)+\cN_{\cC}(\x).
\end{equation}
\end{definition}
Next we establish connections between the dc approximation model \eqref{uL2CP} and the RSMV model \eqref{L2CP}.

\begin{theorem}\label{th:localmin}
Consider $0<t<\min\{1/n,\phi_{\min}/2L_h\}$ with $\displaystyle\phi_{\min}:=\min_{1\leq i\leq  n}\tilde{\phi}_i$. Let $\bar{\x}$ be a lifted stationary point of the dc approximation model \eqref{uL2CP}.
\begin{itemize}
\item[(i)] If there exists $i\in\{1,\ldots,n\}$ such that
$\bar{\x}_i\in(-t,t)$, then $\bar{\x}_i=0$.
\item[(ii)] For $i=1,\ldots,n$, $\bar{d}_i\in\cD(\bar\x_i)$ in \eqref{cond-lift} is unique. That is,
$$
\bar{d}_i=1~\hbox{if }~|\x_i|< t,~\bar{d}_i=2~\hbox{if }~\x_i\geq t,~\hbox{and }~\bar{d}_i=3~\hbox{if }~\x_i\leq- t.
$$
\item[(iii)] ${\bar\x}$ is a local minimizer of model \eqref{L2CP}.
\item[(iv)] If ${\bar\x}$ is a global minimizer of model \eqref{uL2CP}, then it is  a global minimizer of model \eqref{L2CP}.
\end{itemize}
\end{theorem}
Theorem \ref{th:localmin} tells that with a proper choice of parameter \(t\), the lifted stationary point of model \eqref{uL2CP} is at least a local solution of the RSMV model \eqref{L2CP}. Next we develop an efficient algorithm that finds the lifted stationary point of model \eqref{uL2CP}.



\subsection{The proximal dc algorithm}\label{sec:algo}

In this section, we introduce a proximal algorithm to find a lifted stationary point of the dc approximation problem \eqref{uL2CP}. We prove that it has a local linear convergence property by leveraging the Kurdyka-{\L}ojasiewicz (KL) property.
The proximal dc algorithm applied in this paper, a variant of the classic dc algorithm (\calt{tao1997convex}), operates as follows: at each iteration \(k\), it approximates the term \(q_t(\x)\) by its affine minorization and incorporates a proximal term to ensure that the resulting subproblems are well-defined convex problems. For a comprehensive overview of recent theoretical and algorithmic advancements in dc algorithms, we refer to (\calt{le2018dc}).
The KL property is a fundamental tool for proving the convergence rate in Theorem \ref{th-convergence} to be presented. To make the paper self-contained, we present definitions of the KL function and the KL exponent as in (\calt{attouch2010proximal,attouch2013convergence,li2018calculus}).

\begin{definition}\label{def-KL}
 (KL function) The function $f:\R^m\to\R\cup\{+\infty\}$ is said to have the KL property at $\bar{x}\in{\dom}\,\partial f$ if there exist $\eta\in(0,+\infty]$, a neighborhood $\cU$ of $\bar{x}$ and a continuous concave function $\psi:(0,\eta]\to\R_+$ such that
\begin{itemize}[leftmargin=7mm]
  \item[(i)] $\psi(0)=0$ and $\psi$ is continuous differentiable on $(0,\eta)$;
  \item[(ii)] $\psi(s)>0$, for all $s\in(0,\eta]$;
  \item[(iii)] for all $x\in\cU\cap\{x\in\R^n:f(\bar{x})<f(x)<f(\bar{x})+\eta\}$, the KL inequality \(\label{KL-Ieq}
      \psi'(f(x)-f(\bar{x})){\rm dist}(0,\partial f(x))\geq 1
      \) holds.
\end{itemize}
Furthermore, $f$ is said to be a KL function if $f$ satisfies the KL inequality  at each point of ${\rm dom}~\partial f$.
\end{definition}

\begin{definition}(KL exponent) For a proper closed function $f$ satisfying the KL property at $\bar{x}\in{\dom}\,\partial f$, if the corresponding function $\psi$ can be chosen as $\psi(s)=xs^{1-\alpha}$ for some $c>0$ and $\alpha\in[0,1)$, then we say that $f$ has the KL property at $\bar{x}$ with an exponent of $\alpha$.  If $f$ is a KL function and has the same exponent
$\alpha$ at any $\bar{x}\in\dom f$ , then we say that $f$ is a KL function with an exponent of $\alpha$.
\end{definition}

In the following proposition, we show that the essential objective function of model \eqref{uL2CP} is a KL function with an exponent of $1/2$, which ensures the linear convergence of our proximal dc algorithm to be presented in Algorithm 1.
%

\begin{proposition}\label{prop-KL}
The  following essential objective function $f:\R^n\to\R\cup\{+\infty\}$ of problem \eqref{uL2CP} is a KL function with an exponent of $\frac{1}{2}$:
\begin{equation}\label{def-f}
f(\x):=\frac{1}{2}\|W\x\|^2_2+\lambda \|W\x\|_2-\tilde{\mathbf{r}}^T\mathbf{x}+ p_t(\x)-q_t(\x)+\I_{\cC}(\x),
\end{equation}
where $\I_{\cC}(\x)=0$ if $\x\in\cC$ and $\I_{\cC}(\x)=+\infty$ otherwise.
\end{proposition}

The KL property of function $f$ plays an essential role in algorithm design. It provides theoretical assurance regarding the convergence rate of the algorithm. Such theoretical grounding enhances algorithmic reliability and ensures better interpretability of estimates.


For any $\x\in\R^n$, we define the following notation:
$$
\cQ(\x):=\left\{q\in\R^n:\begin{array}{l}
 q_{i}=\phi_i/t~\hbox{if}~\x_i\geq t,\,q_i=-\phi_i/t~\hbox{if}~\x_i\leq -t,\\
~\hbox{and}~  q_i=0,~\hbox{if}~|\x_i|< t,\,\,\, \hbox{for all}\,\, i =1,\ldots,n.
 \end{array}
 \right\}
$$
based on result (ii) in Theorem \ref{th:localmin}. It is obvious that for any $\x\in\R^n$, the singleton set $\cQ(\x)\subseteq \partial q_t(\x)$.
We summarize the proximal dc algorithm in Algorithm \ref{alg-pDCA} and prove its convergence in Theorem \ref{th-convergence}.

%
%

\begin{algorithm}[H]
\caption{Proximal dc algorithm}\label{alg-pDCA}
\vspace{1mm}
Input $\x^0\in \{\x\in\R^n: \e^T\x-1=0\}$ and $\sigma_0>0$. Iterate the following steps for $k=0,1,\dots$:
\vspace{1mm}
\noindent
\begin{description}
 \item[{\sf Step 1.}] Compute $q^k\in \cQ(\x^k)$.
 \item[{\sf Step 2.}] Solve the following optimization problem 
 \begin{equation}\label{alg-sub0}
 \begin{array}{cl}
  \min\limits_{\x} &  g_k(\x):=\frac{1}{2}\|W\x\|^2_2+\lambda \|W\x\|_2-\tilde{\mathbf{r}}^T\mathbf{x}+ p_t(\x)-\langle q^k,\x-\x^k\rangle+\frac{\sigma_k}{2}\|\x-\x^k\|^2_2\\[2mm]
  {\rm s.t.} & \e^T\x-1=0
  \end{array}
\end{equation}
to find $\x^{k+1}$ such that $\e^T\x^{k+1}-1=0$ and $\delta_k\in \partial g_k(\x^{k+1})$ with $\|\delta_k\|_2\leq \frac{\sigma_k}{4}\|\x^{k+1}-\x^{k}\|_2$.
  \item[{\sf Step 3.}] If $\x^{k+1}$ satisfies a preset stopping criterion, terminate; otherwise, update  $\sigma_{k+1}=\gamma_k\sigma_k$ with $\gamma_k>1$.
\end{description}
\end{algorithm}

\begin{theorem}\label{th-convergence}
Assume that the sequence $\{\sigma_k\}$ is convergent. Let $\{\x^k\}$ be the sequence generated by Algorithm \ref{alg-pDCA}. Then  the whole sequence $\{\x^k\}$  converges locally linearly to a lifted stationary point of the dc approximation model \eqref{uL2CP}, and consequently to the local minimizer of the RSMV model \eqref{L2CP}.
\end{theorem}

\subsection{The semismooth Newton-based proximal dc algorithm}
In an attempt to efficiently solve the subproblem \eqref{alg-sub0} in Algorithm \ref{alg-pDCA}, we adapt a specialized semismooth Newton method in \ct{zhao2010newton}. We show that despite the inclusion of the linear constraint, the adapted semismooth Newton method has global convergence with at least a locally superlinear rate. Its fast convergence ensures the overall computational efficiency of the proximal dc algorithm presented in Algorithm \ref{alg-pDCA}.
In this section we provide an overview of the semismooth Newton method while we include detailed explanations in \ref{appendix-ssn}.

The Lagrangian dual associated with the $k$-th subproblem \eqref{alg-sub0}
is given by
\begin{equation}\label{k-dual}
\min\limits_{\y,v}~h_k(\y,v),
\end{equation}
where
\begin{equation}\label{def-h}
\begin{array}{l}
h_k(\y,v):=-\mathcal{M}_{\lambda\|\cdot\|_2}(\y)+\frac{1}{2}\|\y\|^2_2+v\\[2mm]
~~~~~~~~~~~~~~-\sigma_k\mathcal{M}_{p_t/\sigma_k}\big(\x^k-(W^T\y+\e v-Q_k)/\sigma_k\big)
+\frac{\sigma_k}{2}\|x_k-(W^T\y+\e
v-Q_k)/\sigma_k\|^2_2.
\end{array}
\end{equation}
Here $\cM_{\varphi}(z):=\min_x\{\varphi(x)+\frac{1}{2}\|x-z\|^2_2\}$ is the Moreau-Yosida regularization associated with the proper closed convex function $\varphi:\R^n\to\R$ at $z\in\R^n$. Let $\prox_{\varphi}(z):=\arg\min_x \{\varphi(x)+\frac{1}{2}\|x-z\|^2_2\}$ be the proximal mapping of $\varphi$ at $z\in\R^n$. It follows from Theorem 2.26 of \ct{rockafellar2009variational} that the function $\cM_{\varphi}(\cdot)$ is smooth with Lipschitz continuous gradient
$\nabla\cM_{\varphi}(z)=z-\prox_{\varphi}(z)$.
 Consequently, the function $h_k$ is convex and smooth with Lipschitz continuous gradient:
$$
 \nabla h_k(\y,v) = \left(\begin{array}{c}
 \prox_{\lambda\|\cdot\|_2}(\y)-W\prox_{p_t/\sigma_k}(\tilde{x}_k(\y,v))\\[2mm]
 -\e^T\prox_{p_t/\sigma_k}(\tilde{x}_k(\y,v))+1
 \end{array}\right),
 $$
where $\tilde{x}_k(\y,v):=\x^k-(W^T\y+\e v-Q_k)/\sigma_k$. The first optimality condition of problem \eqref{k-dual} implies that its optimal solution can be obtained by solving the linear system:
\begin{equation}\label{ssn-eq}
\nabla h_k(\y,v)=0.
\end{equation}
We further define a multifunction $\mathcal{G}_k:\R^{n+1}\rightrightarrows\mathbb{S}^{n+1}$ to characterize the second-order information of $h_k$:
\begin{equation}\label{def-G}
\mathcal{G}_k(\y,v)=\left\{\left(\begin{array}{cc}
U+\sigma^{-1}_k WVW^T & \sigma^{-1}_kWV\e\\[2mm]
\sigma^{-1}_k\e^TVW^T &  \sigma^{-1}_k\e^TV\e
\end{array}
\right):
U\in\partial_B\prox_{\lambda\|\cdot\|_2}(\y),\,\,V\in \partial_B\prox_{p_t/\sigma_k}(\tilde{x}^k(\y,v)) \right\}.
\end{equation}
We show in \ref{appendix-th3} that the gradient $\nabla h_k$ is strongly semismooth  with respect to $\cG_k$, any element in $\cG_k(\y,v)$ is positive semi-definite, and all the elements in $\cG_k(\y,v)$ at the solution to problem \eqref{ssn-eq} are positive definite. With these findings, we can modify the semismooth Newton method discussed in \ct{zhao2010newton} for addressing the subproblem \eqref{alg-sub0} of the proximal algorithm and prove its global convergence with a minimum locally superlinear convergence rate in Theorem \ref{prop-SN}.

\begin{theorem}\label{prop-SN}
Let $\{(\y^{k,j},v^{k,j})\}$ be the sequence generated by Algorithm \ref{alg-SNpDCA}.
Then  $\{(\y^{k,j},v^{k,j})\}$ is
well-defined and converges  to the solution $(\y^{k,*},v^{k,*})$. Moreover, the local convergence rate is at least superlinear:
$$
\|(\y^{k,j+1},v^{k,j+1}) - (\y^{k,*},v^{k,*})\| = O(\|(\y^{k,j},v^{k,j}) - (\y^{k,*},v^{k,*})\|^{1+\tau}),
$$
where $\tau\in (0,1]$ is the parameter given in Algorithm \ref{alg-SNpDCA}.
\end{theorem}

In Algorithm \ref{alg-SNpDCA}, we succinctly encapsulate our complete solution scheme for the RSMV model: the Semismooth Newton-based Proximal DC Algorithm (SN-pDCA). The SN-pDCA expands the proximal dc algorithm (i.e. Algorithm \ref{alg-pDCA}) by including the adapted semismooth Newton method in Step 2 for solving subproblems.

\begin{algorithm}
  Initialize {$\x^0\in \{\x\in\R^n: \e^T\x-1=0\}$, $\sigma_0>0$, $k=0$}
  \;
  \While {$\x^{k}$ dose not satisfies a preset stopping criterion}{
    $q^k\in \cQ(\x^k)$ \;
    Select $\mu \in (0,1/2)$, $\bar{\eta} \in (0,1)$, $\tau \in (0,1]$, $\tau_1,\tau_2\in(0,1)$, $\beta \in (0,1)$, $\y^{k,0}\in\mathbb{R}^m,\, v^{k,0}\in\R$, $j=0$\;

    \Repeat{$\x^{k+1} = \prox_{p_t/\sigma_k}\big(\x^k-(W^T\y^{k,j+1}+\e v^{k,j+1}-q^k)/\sigma_k\big)$, $\x^{k+1}=\x^{k+1}/\sum^n_{i=1}\x^{k+1}_i$ satisfies $\delta_k\in \partial g_k(\x^{k+1})$ with $\|\delta_k\|_2\leq \frac{\sigma_k}{4}\|\x^{k+1}-\x^{k}\|_2$}{
      \begin{description}
\item[\sf S1.] (Newton Direction) Choose $G_{k,j}\in \mathcal{G}_k(\y^{k,j},v^{k,j})$.
 Solve the following linear system
$$
(G_{k,j}+\epsilon_j I) d = -\nabla h_k(\y^{k,j},v^{k,j}),\,\,\,\epsilon_j:=\tau_1\min\{\tau_2,\|\nabla h_k(\y^{k,j},v^{k,j})\|_2\},
$$
 by the practical conjugate gradient algorithm   to find $d_{k,j}$ such that $$\|(G_{k,j}+\epsilon_j I) \red{d} +\nabla h_k(\y^{k,j},v^{k,j}) \|_2 \leq \min(\bar{\eta},\|\nabla h_k(\y^{k,j},v^{k,j})\|^{1+\tau}_2).$$
\item[\sf S2.] (Line Search) Set $ \alpha_j = \beta^{m_j}$, where $m_j$ is the smallest nonnegative integer $m$ for which
\begin{equation*}
 h_k\big((\y^{k,j},v^{k,j}) + \beta^m d_{k,j}\big) \leq  h_k(\y^{k,j},v_{k,j}) + \mu\beta^m \langle \nabla h_k(\y^{k,j},v^{k,j}),d_{k,j} \rangle.
\end{equation*}
\item[\sf S3.] $ (\y^{k,j+1},v^{k,j+1}) = (\y^{k,j},v^{k,j}) + \alpha_j d_{k,j} $ and $j\leftarrow j+1$.	
\end{description}
    }
    Update $\sigma_{k+1}=\gamma_k\sigma_k$ with $\gamma_k>1$\;
    Set $k\leftarrow k+1$\;
  }
  \caption{SN-pDCA}\label{alg-SNpDCA}
\end{algorithm}


\section{\blue{Numerical Results}}\label{sec:BB_comp}
In this section, we adopt the SN-pDCA introduced in Section 3 to obtain RSMV portfolios numerically. We evaluate performance of the SN-pDCA by examining its computational scalability and estimating quality of the RSMV portfolios it generated. Moreover, we illustrate properties of the RSMV portfolios. Proposition \ref{conds} tells that with a simplified covariance matrix, increasing the uncertainty level can first decrease the number of traded assets and encourage diversification subsequently. In this section, we show computationally that the property regarding the cardinality surface of the RSMV portfolio still holds under a general covariance matrix. All our computational results are obtained by running MATLAB 2018b
on a Windows 10 laptop equipped with an i7-10510U CPU @ 1.80GHz 2.30 GHz and 32 GB memory. 



\subsection{Solution quality and computational scalability}
%



This section evaluates performance of the SN-pDCA for solving the RSMV model \eqref{L2CP}. Note that exact solutions of small-size RSMV models can be obtained using CPLEX. Thus, we begin with small-size examples and estimate quality of the SN-pDCA solutions by comparing them with CPLEX solutions. For large-scale examples that cannot be solved using CPLEX, we compute solutions of L1MV model \eqref{L1MV} as benchmark, which are suboptimal solutions to the RSMV model \eqref{L2CP} and are referred to as L1MV solutions in the subsequent analysis. Specifically, we obtain the L1MV solutions through the application of the semismooth Newton-based proximal point algorithm (SN-PPA), which constitutes a modification of Algorithm \ref{alg-pDCA}. \footnote{We refer to the solution of the L1MV model \eqref{L1MV} as the L1MV solution instead of the SN-PPA solution since the L1MV model \eqref{L1MV} is strictly convex and its unique solution does not rely on the numerical approach SN-PPA.} Additionally, drawing from our numerical experience in the domain of dc programming, we employ the L1MV solution as the initial starting point for the SN-pDCA.

We generate RSMV examples using monthly data from the {\it Fama-French Data Library} (FF), the {\it Standard $\&$ Poor’s 500 stocks} (SPX) and the {\it Russell 2000 stocks} (RUT) for estimating the mean and the covariance matrix. Details of the datasets are provided in Table \ref{dataset}. In particular, ``FFInd" represents ``Industry Portfolios" and FF100 stands for ``100 Portfolios Formed on Size and Investment". For SPX and RUT, we include constituents that are present throughout the entire sample period being considered. As a consequence, there are 326 assets in the SPX dataset and 1074 assets in the RUT dataset. Datasets RUT500 and RUT800 are constructed by randomly selecting 500 and 800 assets from the RUT dataset, respectively.

\begin{figure}[t]
    \centering
    \includegraphics[width=\textwidth]{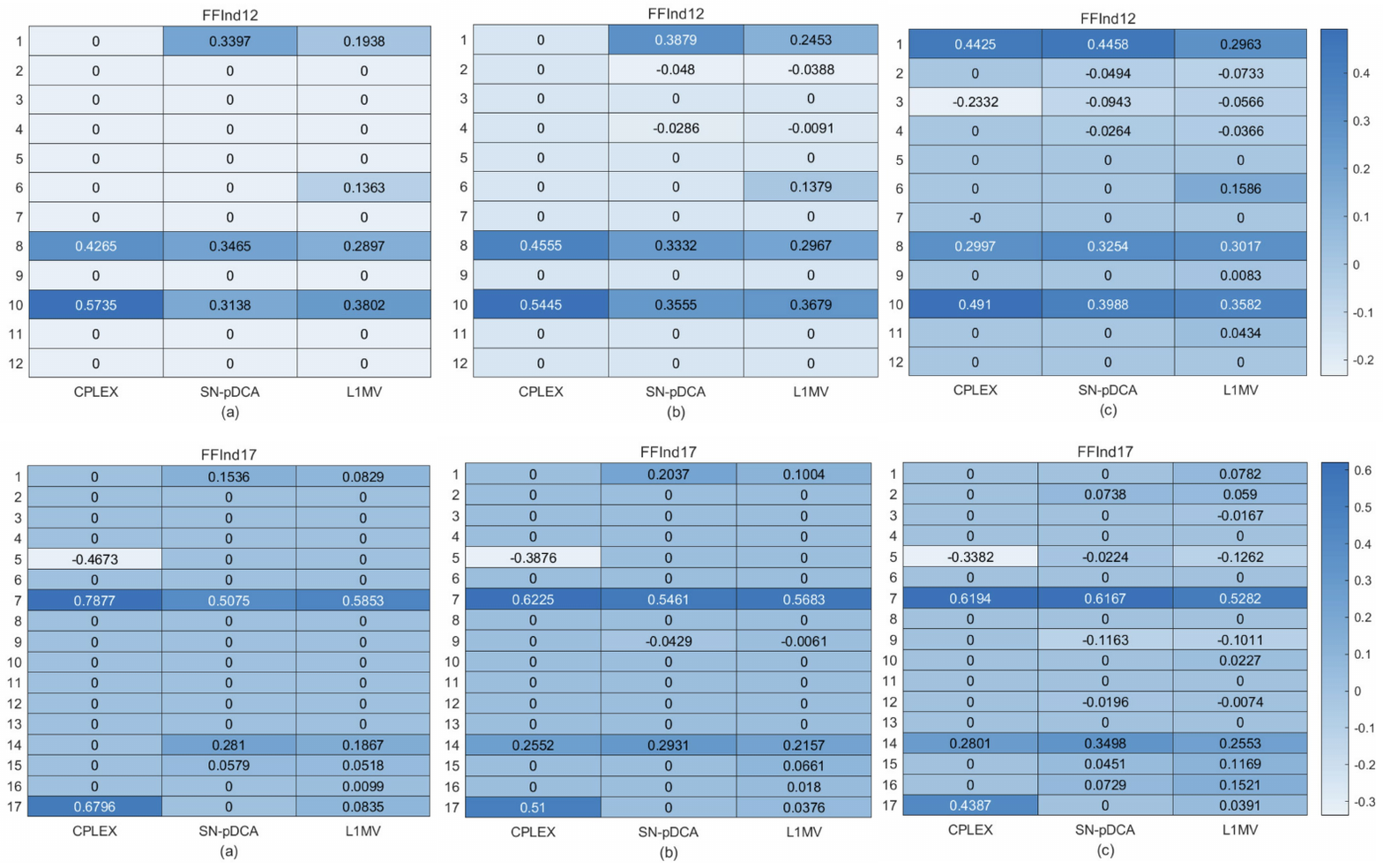}
    \vspace{-2cm}
     \caption{The heatmaps of the CPLEX, SN-pDCA, and L1MV solutions generated using datasets FFInd12 and FFInd17. Model parameters are $\kappa=1$, \(\bm{\phi}=\phi\bm{e}\) with $\phi=10^{-3}$, \(\epsilon=0.1\) in (a),  \(\epsilon=1\) in (b), and \(\epsilon=10\) in (c).}\label{fig-heatmap}
 \end{figure}

\begin{table}[H]
\centering
\begin{tabular}{lccc|lccc}
\hline
Dataset& \tabincell{c}{Number\\ of assets} & Sample Period & Frequency & Dataset& \tabincell{c}{Number\\ of assets}  & Sample Period & Frequency\\
FFInd12 & 12 & 01/2013-12/2022 & Monthly&SPX326 & 326 &01/2013-12/2022 & Daily\\
FFInd17 & 17  & 01/2013-12/2022 & Monthly&RUT500&500 &01/2014-12/2018 & Daily\\
FFInd30 & 30 &01/2013-12/2022 & Monthly&RUT800&800 &01/2014-12/2018 & Daily\\
FFInd48&48& 01/2013-12/2022 & Monthly &RUT1074&1074 &01/2014-12/2018 & Daily\\
FF100 &100&01/2013-12/2022& Monthly&&& &\\
\hline
\end{tabular}\\
\caption {Historical return datasets.}\label{dataset}
\end{table}
As CPLEX can only find the exact RSMV portfolio when the dimension is relatively small, we first compute optimal portfolio weights using datasets FFInd12 and FFInd17 under different values of \(\epsilon\) and present the results in Figure \ref{fig-heatmap}. We terminate the CPLEX solver when either the default stopping criterion is met at $10^{-3}$ or when the computation time reaches 600 seconds. We stop the SN-pDCA and the SN-PPA for the L1MV when $\|x^{k+1}-x^k\|/(1+\|x^k\|)\leq 10^{-5}$.

It can be seen from Figure \ref{fig-heatmap} that the SN-pDCA solution is a better approximate to the CPLEX solution, which is the exact RSMV portfolio, compared to the L1MV solution. In addition, the index set of nonzeros in the SN-pDCA solution is a subset of that in the L1MV solution. As a consequence, we are motivated to expedite the SN-pDCA by reducing the dimension of model \eqref{uL2CP} based on the index set of nonzeros in the solution to the L1MV model \eqref{L1MV} with the relative error of iterations reaches $10^{-3}$. We refer to this accelerated version as Ac-SN-pDCA, which as the SN-pDCA is also terminated when the relative error of iterations reaches $10^{-5}$.

In Table \ref{table-com2Cplex} we evaluate performance of different computational methods using all the datasets listed in Table \ref{dataset}.  The evaluation criteria include the following metrics: the objective value of the RSMV model \eqref{L2CP}, the cardinality (i.e., number of nonzeros) of the optimal/suboptimal RSMV portfolio, and the computational time.
 Table \ref{table-com2Cplex} clearly indicates a significant increase in computational time for CPLEX as the dimension exceeds 48. For the first four datasets with no more than 48 assets,  CPLEX solutions are exact and can be used to compute the relative errors of SN-pDCA, Ac-SN-pDCA, and L1MV solutions. According to Table \ref{table-com2Cplex},  the relative errors of SN-pDCA and Ac-SN-pDCA solutions are less than 10\% and are obviously lower than those of the L1MV solutions.

As dimensions increase, CPLEX struggles to attain a superior objective value within the given computational constraints. In contrast, both SN-pDCA and AC-SN-pDCA demonstrate the ability to effectively reduce the dimensionality within a reasonable computational time frame and to obtain suboptimal portfolios better than the L1MV portfolios in terms of the objective value and portfolio cardinality. Additionally, we observe that compared to the SN-pDCA, its accelerated version (i.e., AC-SN-pDCA) delivers superior performance in less time, particularly in higher dimensions. This improvement may be attributed to reduced computational errors. However, it's essential to note that while the accelerated version is based on numerical insights, further theoretical validation is warranted.


\begin{table}[H]
\centering
{\small
\begin{tabular}{c|cccc|cccc|cccc}
\hline
\multirow{2}{*}{Dataset}& \multicolumn{4}{c|}{Objective value}  & \multicolumn{4}{c|}{Portfolio cardinality} & \multicolumn{4}{c}{Computational time (seconds)}     \\ \cline{2-13}
 & \multicolumn{1}{c}{CPLEX}& \tabincell{c}{SN-\\ pDCA}  & \tabincell{c}{Ac-SN-\\ pDCA} & L1MV   & \multicolumn{1}{c}{CPLEX}& \tabincell{c}{SN-\\ pDCA}  & \tabincell{c}{Ac-SN-\\ pDCA} & L1MV  & \multicolumn{1}{c}{CPLEX}& \tabincell{c}{SN-\\ pDCA}  & \tabincell{c}{Ac-SN-\\ pDCA} & L1MV      \\ \hline
FFInd12 &  {\bf 0.0385} & 0.0398 & 0.0398 & 0.0413
&  4 & 5 & 5 & 6  & 0.17 & 0.11 & 0.11 & 0.11\\ \hline
FFInd17 &  {\bf 0.0366} & 0.0379 & 0.0379 & 0.0400
&  2 & 4 & 4 &6 &  0.26 & 0.13  & 0.15 & 0.11\\ \hline
FFInd30 &  {\bf 0.0374} & 0.0390 & 0.386 & 0.0396
&  4 & 5 & 4 &5  &  0.52 & 0.15 & 0.15 & 0.15\\ \hline
FFInd48 & {\bf 0.0420} & 0.0451  & 0.0451 & 0.0509
& 4 & 9 & 9 & 15 & 7.81 & 0.11 & 0.11 & 0.15 \\\hline
FF100 & {\bf 0.0458} & 0.0504 & 0.0504 & 0.0564
& 5 & 8 & 8 & 14 & 600.00 & 0.29 & 0.30 & 0.30\\\hline
SPX326 & 0.0405 & 0.0404 & {\bf0.0401} & 0.0646
& 5 & 13 & 12 & 38 & 600.00 & 1.97 & 1.80 & 1.97 \\\hline
RUT500 & 0.0919 & 0.0821& {\bf 0.0797} & 0.0981
& 8 & 23 & 16 & 35 & 600.00 & 1.84 & 1.70 & 1.60 \\\hline
RUT800 & 0.1581 & 0.0848 & {\bf 0.0825} & 0.1021
& 2 & 23 & 19 & 39 & 600.00 & 4.62 & 3.53 & 4.19\\\hline
RUT1074 & 7335.10 & {\bf 0.0998} & {\bf 0.0998} & 0.1268
& 1 & 24 & 24 & 47 & 600.00 & 9.27 & 8.80 & 8.10\\\hline
\end{tabular}}
\caption{Computational performance of CPLEX, SN-pDCA, Ac-SN-pDCA, and L1MV.
The model parameters are
$\kappa=1,\,\varepsilon=1,$ and \(\bm{\phi}=\phi\bm{e}\) with $\phi=10^{-3}$. }\label{table-com2Cplex}
\end{table}

\subsection{Cardinality surface}\label{Cardinality surface}

We plot the cardinality surface of the RSMV portfolio using  four different datasets described in Table \ref{dataset} that include return data between the given dates to estimate the mean and covariance matrix of the returns. For datasets FFInd17 and FFInd30, we use CPLEX to obtain the global optimal solution of the RSMV model \eqref{rsmv}. 
For large-scale datasets SPX326 and RUT500, we implement the SN-pDCA to obtain an approximate RSMV portfolio.

\begin{figure}
\centering
\begin{subfigure}{.6\textwidth}

  \includegraphics[width=.7\linewidth]{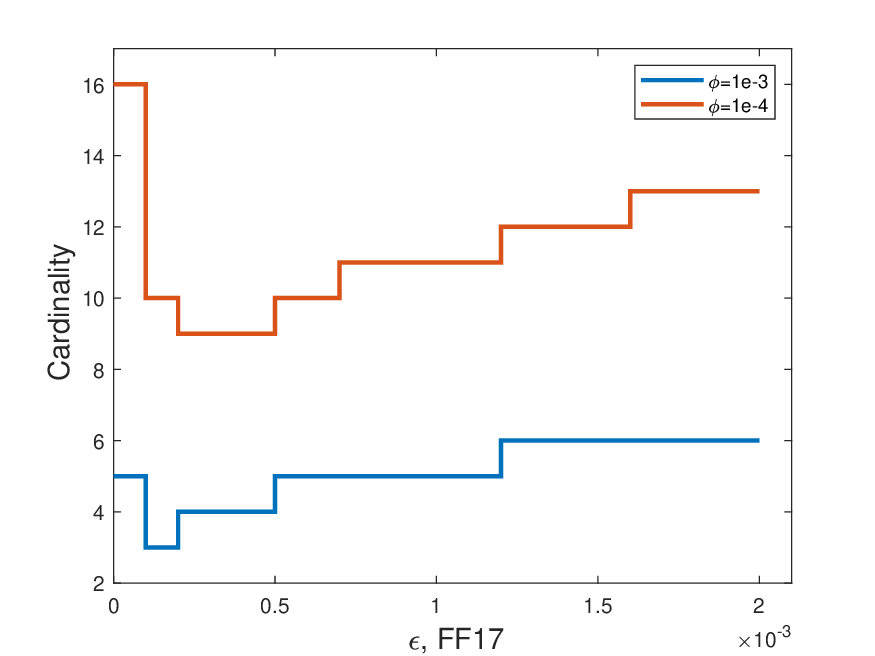}
  \label{fig:sub1}
\end{subfigure}%
\begin{subfigure}{.6\textwidth}

  \includegraphics[width=.7\linewidth]{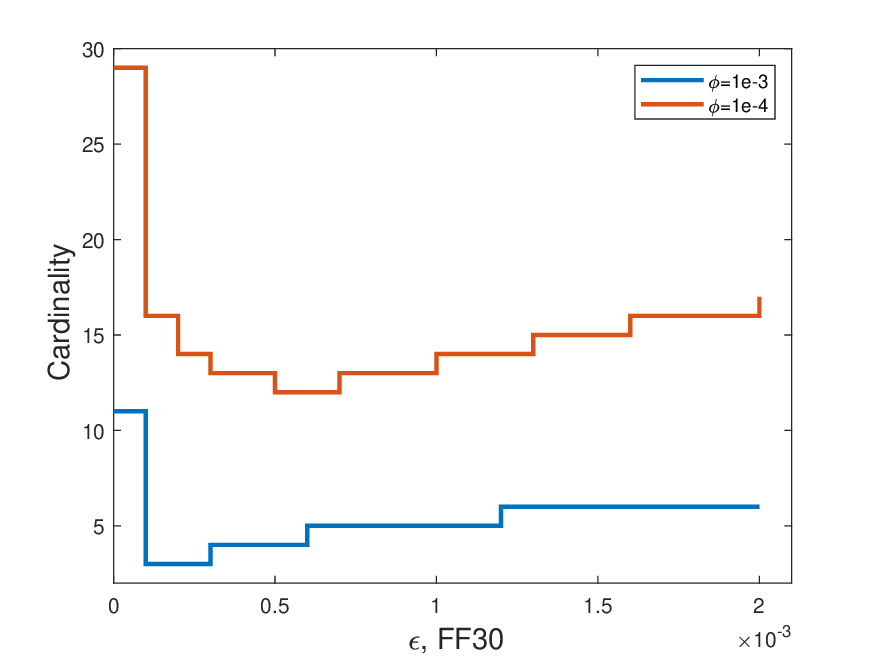}
  \label{fig:sub2}
\end{subfigure}
\label{fig:test}
\begin{subfigure}{.6\textwidth}
  \includegraphics[width=.7\linewidth]{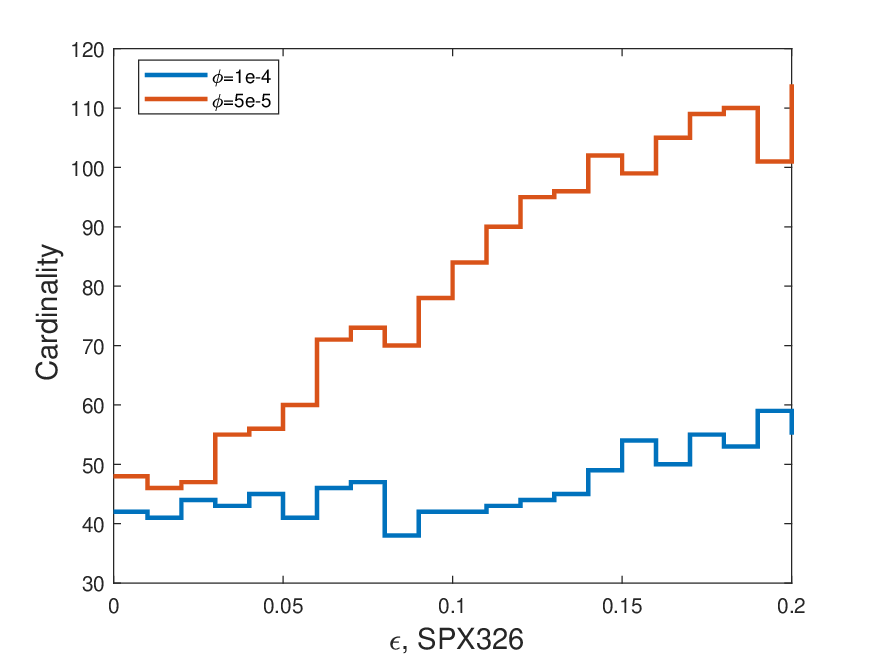}
  \label{fig:sub3}
\end{subfigure}%
\begin{subfigure}{.6\textwidth}
  \includegraphics[width=.7\linewidth]{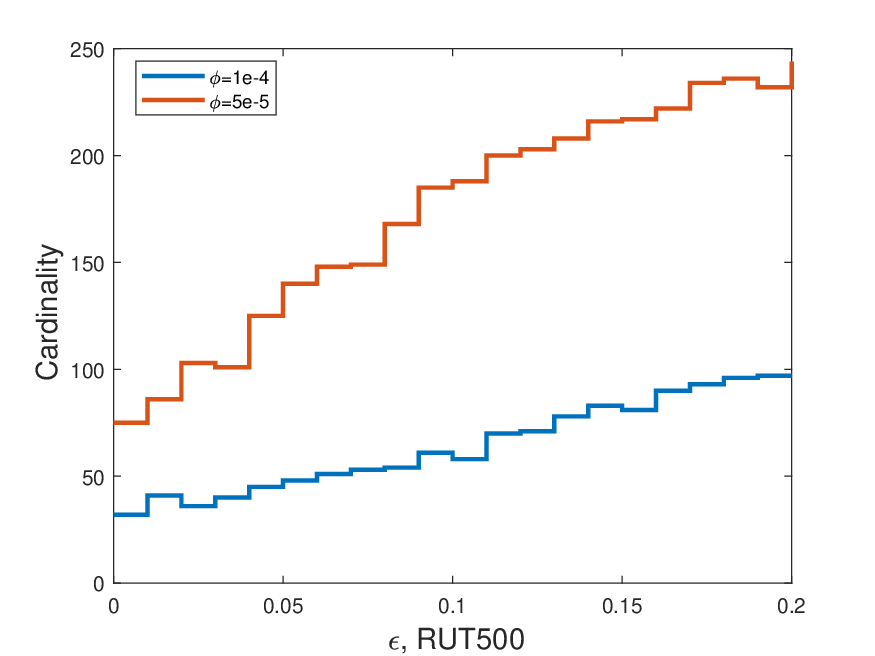}
  \label{fig:sub4}
\end{subfigure}
\caption{Cardinalities of the RSMV portfolios for FF17, FF30, SPX326, and RUT500 datasets.}\label{fig:cardS}
\end{figure}

\blue{We arbitrarily set $\kappa=1$. For a given $\phi$, we plot the cardinality curve in terms of the uncertainty level $\epsilon$ in Figure \ref{fig:cardS}. We let \(\epsilon\) range from $0$ to $0.002$ with an incremental step size $0.0001$ for datasets FFInd17 and FFInd30. For datasets SPX326 and RUT500}, we vary \(\epsilon\) vary from $0$ to $0.2$ with an incremental step size $0.01$ It can be clearly seen that for the class of portfolios being considered, under a given fixed transaction cost, the cardinality of the portfolio could decrease, remain the same, or increase as the uncertainty level increases. It verifies that the insights obtained in Section \ref{sec:SimBnd} apply to a general setting of covariance matrix. 

\section {Conclusions}\label{sec:Conc}
In this paper, we extend the classical MV framework and propose a robust and sparse portfolio selection model, which mathematically is a cardinality constrained quadratic program and is challenging to solve under high dimension. We develop an efficient semismooth Newton based proximal dc algorithm that finds a global or local solution, and prove its superlinear local convergence rate. Moreover, we provide a fundamental understanding of the impact of parameter uncertainty and fixed transaction costs on the portfolio cardinality. {Specifically, we show that the robust MV portfolio is indeed a unified framework that can generalize a set of well-studied portfolios. We also characterize the conditions, both theoretically and numerically, under which the parameter uncertainty could promote or discourage diversification, unveiling the diversification paradox.}

\section*{Acknowledgement}
Jingnan Chen's research was partially supported by the National Natural Science Foundation of China (NSFC) under grant 72171012 and the Fundamental Research Funds for the Central Universities. Ning Zhang's research was partially supported by the National Natural Science Foundation of China (NSFC) under grant 12271095.

\bibliographystyle{unsrtnat}

\bibliography{InPortfolio}

\newpage
\begin{appendix}
\begin{center}
	\section*{ {APPENDICES: ``Robust and Sparse Portfolio Selection: Quantitative Insights and Efficient Algorithms'' \\
	}}
\end{center}	
\vskip .1in

\section{Proof of Proposition \ref{thrm:VVR}}
The RMV model is equivalent to the following problem
\begin{equation*}
\displaystyle\max_{\lambda, \mathbf{M}\succeq\mathbf{0}}\min_{t, \mathbf{x}}\kappa\,\mathbf{x}^T\bm{\Sigma}\mathbf{x}-\left(\bar{\mathbf{r}}+\lambda\mathbf{e}+2\mathbf{y}\right)^T\mathbf{x}+\left(\sqrt{\varepsilon}-\left<\mathbf{V}, \bm{\Sigma}^{-1}\right>-v\right)t+\lambda,
\end{equation*}
where $\lambda$ and $\mathbf{M}=\left[\begin{array}{cc}\mathbf{V}&\mathbf{y}\\\mathbf{y}^T&v\end{array}\right]$ are the dual variables. By solving the inner minimization problem, it reduces to
\begin{equation}\label{eq:dualvvar}
\begin{array}{cl}
\displaystyle\max_{\lambda, \mathbf{M}} & \lambda-\frac{1}{4\kappa}\left(\bar{\mathbf{r}}+\lambda\mathbf{e}+2\mathbf{y}\right)^T\bm{\Sigma}^{-1}\left(\bar{\mathbf{r}}+\lambda\mathbf{e}+2\mathbf{y}\right)\\
\text{subject to}& \displaystyle\sqrt{\varepsilon}-\left<\mathbf{V}, \bm{\Sigma}^{-1}\right>-v= 0,  \mathbf{M}\succeq \mathbf{0} , \lambda\in\mathscr{R}.
\end{array}
\end{equation}
Note that, for given $\mathbf{y}$, the optimal $\lambda^\star(\mathbf{y})$ is obtained by solving the following quadratic program
\begin{equation*}
\displaystyle\max_{\lambda\in\mathscr{R}}-\frac{\mathbf{e}^T\bm{\Sigma}^{-1}\mathbf{e}}{4\kappa}\lambda^2+\left(1-\frac{\mathbf{e}^T\bm{\Sigma}^{-1}\left(\bar{\mathbf{r}}+2\mathbf{y}\right)}{2\kappa}\right)\lambda
\end{equation*}
with $\displaystyle\lambda^\star(\mathbf{y})$ being $\frac{2\kappa-\mathbf{e}^T\bm{\Sigma}^{-1}(\bar{\mathbf{r}}+2\mathbf{y})}{\mathbf{e}^T\bm{\Sigma}^{-1}\mathbf{e}}$. Then, substituting $\lambda^\star(\mathbf{y})$ into (\ref{eq:dualvvar}), it becomes
\begin{equation*}
\begin{array}{cl}
\displaystyle\max_{\mathbf{M}} & v_\text{MV}+\widehat{\mathbf{r}}^T\mathbf{y}-\mathbf{y}^T\widehat{\Sigma}\mathbf{y}\\
\text{subject to}& \sqrt{\varepsilon}-\left<\mathbf{V}, \bm{\Sigma}^{-1}\right>-v= 0,\mathbf{M}\succeq \mathbf{0},
\end{array}
\end{equation*}
where $\widehat{\mathbf{r}}=-2\,\mathbf{x}_{\text{MV}}$, $\widehat{\bm{\Sigma}}=\frac{1}{\kappa}\left(\bm{\Sigma}^{-1}-\frac{\bm{\Sigma}^{-1}\mathbf{e}\mathbf{e}^T\bm{\Sigma}^{-1}}{\mathbf{e}^T\bm{\Sigma}^{-1}\mathbf{e}}\right)$, and it is a convex problem since $\widehat{\bm{\Sigma}}\succeq\mathbf{0}$ is a semidefinite matrix. From Schur complement, we have $\mathbf{M}\succeq \mathbf{0}\Leftrightarrow\frac{1}{v}\mathbf{y}\mathbf{y}^T\preceq\mathbf{V}, v>0$, and substituting this into the equality constraint $\sqrt{\varepsilon}-\left<\mathbf{V}, \bm{\Sigma}^{-1}\right>-v= 0$, it becomes $\mathbf{y}^T\bm{\Sigma}^{-1}\mathbf{y}\leq v(\sqrt{\varepsilon}-v)$. Since (\ref{eq:dualvvar}) is a maximization problem, we set $v=\sqrt{\varepsilon}/2$ which maximizes the function $v(\sqrt{\varepsilon}-v)$ to obtain
\begin{equation}\label{dmvo}
v_\emph{MV}+\underbrace{ \displaystyle\max_{\mathbf{y}^T\bm{\Sigma}^{-1}\mathbf{y}\leq {\varepsilon}/{4}}  - \mathbf{y}^T\widehat{\bm{\Sigma}}\mathbf{y}-2\mathbf{x}_{\emph{MV}}^T\mathbf{y}}_{\epsilon\emph{-induced QCQP}}
\end{equation}


The KKT optimality conditions of (\ref{dmvo}) are given by
\begin{equation*}
\left\{
\begin{array}{l}\widehat{\mathbf{r}}=2\left(\widehat{\bm{\Sigma}}+\rho\bm{\Sigma}^{-1}\right)\mathbf{y},\\
\mathbf{y}^T\bm{\Sigma}^{-1}\mathbf{y}=\varepsilon/4,
\end{array}\right.
\end{equation*}
where $\rho>0$ is the dual variable associated with the quadratic constraint $\mathbf{y}^T\bm{\Sigma}^{-1}\mathbf{y}\leq\varepsilon/4$. Hence, for given $\rho>0$, the optimal $\mathbf{y}(\rho)=\frac{1}{2}\left(\widehat{\bm{\Sigma}}+\rho\bm{\Sigma}^{-1}\right)^{-1}\widehat{\mathbf{r}}$ with $\left(\widehat{\bm{\Sigma}}+\rho\bm{\Sigma}^{-1}\right)^{-1}=\kappa\,\left(\frac{\bm{\Sigma}}{1+\kappa\rho}+\frac{\mathbf{e}\mathbf{e}^T}{(\kappa\rho)(1+\kappa\rho)\mathbf{e^T}\bm{\Sigma}^{-1}\mathbf{e}}\right)
$. Plugging this into the function $f(\rho)=\mathbf{y}(\rho)^{T}\bm{\Sigma}^{-1}\mathbf{y}(\rho)$, we obtain
$
f(\rho)=\frac{\kappa^2}{4(1+\kappa\rho)^2}\left(\widehat{\mathbf{r}}^T\bm{\Sigma}\widehat{\mathbf{r}}+\frac{2(\widehat{\mathbf{r}}^T\mathbf{e})^2}{\kappa\rho\,\mathbf{e}^T\bm{\Sigma}^{-1}\mathbf{e}}+\frac{(\widehat{\mathbf{r}}^T\mathbf{e})^2}{(\kappa\rho)^2\,\mathbf{e}^T\bm{\Sigma}^{-1}\mathbf{e}}\right)
$
with a more compact form as $
f(\rho)=\frac{\kappa^2}{4(1+\kappa\rho)^2}\norm{\mathbf{L}^T\widehat{\mathbf{r}}+\frac{(\widehat{\mathbf{r}}^T\mathbf{e})\mathbf{L}^{-1}\mathbf{e}}{\kappa\rho\, \mathbf{e}\bm{\Sigma}^{-1}\mathbf{e}}}^2$,
where $\bm{\Sigma}=\mathbf{L}\mathbf{L}^T$ and $\mathbf{L}$ is a  lower triangular matrix from Cholesky decomposition. It is straightforward to check that $f(\rho)$ is a monotone decreasing function with $\lim_{\rho\rightarrow 0}f(\rho)= \infty$ and $\lim_{\rho\rightarrow +\infty}f(\rho)= 0$. Hence, there always exists a unique $\rho^\star>0$ satisfying the equation $f(\rho^\star)=\varepsilon/4$, which implies the quadratic constraint is active at optimal solution. Plugging $\mathbf{y}(\rho^\star)$ and $\lambda^{\star}(\mathbf{y}(\rho^{\star}))$ into  $\mathbf{x}_{\text{RMV}}=\frac{1}{2\kappa}\bm{\Sigma}^{-1}\left(\bar{r}+\lambda^{\star}(\mathbf{m}(\rho^{\star}))\mathbf{e}+2\mathbf{m}(\rho^\star)\right)$ and $ \mathbf{y}(\rho^\star)^T\widehat{\bm{\Sigma}}\mathbf{y}(\rho^\star)+2\mathbf{x}_{\text{MV}}^T\mathbf{y}(\rho^\star)$, we can establish the result in Proposition \ref{thrm:VVR}.

\section{Proof of Proposition \ref{wwvar}}
The dual of (\ref{wvar}) is given by
\begin{equation*}
\begin{array}{cl}
\displaystyle\max_{\lambda, \mathbf{M}} & \lambda\\
\text{subject to}& \sqrt{\varepsilon}-\left<\mathbf{V},\bm{\Sigma}^{-1}\right>-v=0,  \lambda\mathbf{e}+2\mathbf{y}+\bar{\mathbf{r}}=0, \mathbf{M}\succeq\mathbf{0},
\end{array}
\end{equation*}
where $\lambda$ and $\mathbf{M}$ are dual variables. We can show that the optimal $\lambda^\star=\frac{-\bar{\mathbf{r}}^T\bm{\Sigma}^{-1}\mathbf{e}+\sqrt{\left(\bar{\mathbf{r}}^T\bm{\Sigma}^{-1}\mathbf{e}\right)^2-\mathbf{e}^T\bm{\Sigma}^{-1}\mathbf{e}\left(\bar{\mathbf{r}}^T\bm{\Sigma}^{-1}\bar{\mathbf{r}}-\varepsilon\right)}}{\mathbf{e}^T\bm{\Sigma}^{-1}\mathbf{e}}$ by a similar derivation as in the proof of Theorem 1. Hence, we can obtain the worst-case VaR portfolio $\mathbf{x}_{\text{WVaR}}$ by solving the following KKT optimality conditions:
\begin{equation*}
\left[\begin{array}{cc}\frac{2\mathbf{y}\mathbf{y}^T}{\sqrt{\varepsilon}}&\mathbf{y}\\\mathbf{y}^T&\frac{\sqrt{\varepsilon}}{2}\end{array}\right]\left[\begin{array}{cc}t\bm{\Sigma}^{-1}&\mathbf{x}\\\mathbf{x}^T&t\end{array}\right]=0, \mathbf{e}^T\mathbf{x}=1,
\end{equation*}
where $\mathbf{y}=-\left(\bar{\mathbf{r}}+ \lambda^\star\mathbf{e}\right)/2$.

\section{Proof of Proposition 3}
Denote $\bm{\Sigma}=\mathbf{U}\bm{\Lambda}\mathbf{U}^T$ as the eigenvalue decomposition of $\bm{\Sigma}$, $\widetilde{\bm{\Sigma}}^{-1}=\mathbf{U}\widetilde{\bm{\Lambda}}^{-1}\mathbf{U}^T$ with  $\widetilde{\bm{\Lambda}}=\bm{\Lambda}+\left(\sqrt{\varepsilon}/{\kappa}\right)\mathbf{I}$, and $\mathbf{u}=\mathbf{U}\mathbf{e}$. The Euclidean distance between $\mathbf{x}_\text{MV}^{l_2}$ and $\mathbf{x}_\text{EW}$ is bounded by
\begin{equation*}\label{dist}
\begin{array}{lll}
\displaystyle\norm{\mathbf{x}_\text{MV}^{l_2}-\mathbf{x}_\text{EW}}_2&\leq&\displaystyle\frac{\norm{\bar{\mathbf{r}}}_2}{2\kappa}\norm{\widetilde{\bm{\Lambda}}^{-1}-\frac{\widetilde{\bm{\Lambda}}^{-1}\mathbf{u}\mathbf{u}^T\widetilde{\bm{\Lambda}}^{-1}}{\mathbf{u}^T\widetilde{\bm{\Lambda}}^{-1}\mathbf{u}}}_2+\sqrt{N}\norm{\frac{\widetilde{\bm{\Lambda}}^{-1}}{\mathbf{u}^T\widetilde{\bm{\Lambda}}^{-1}\mathbf{u}}-\frac{1}{N}\mathbf{I}}_2\\
&\leq&\displaystyle\frac{\norm{\bar{\mathbf{r}}}_2}{2\kappa}\frac{1}{\lambda_{[N]}+\sqrt{\varepsilon}/\kappa}+\sqrt{N}\left( \frac{(\lambda_{[N]}+\sqrt{\varepsilon}/\kappa)^{-1}}{\sum_{i=1}^N u_i^2(\lambda_i+\sqrt{\varepsilon}/\kappa)^{-1}}-\frac{1}{N}\right) \\
&=&\displaystyle\frac{\norm{\bar{\mathbf{r}}}_2}{2\kappa}\frac{1}{\lambda_{[N]}+\sqrt{\varepsilon}/\kappa}+ \frac{\sum_{i=1}^{N}(\lambda_i-\lambda_{[N]})(\lambda_{i}+\sqrt{\varepsilon}/\kappa)^{-1}(\lambda_{[N]}+\sqrt{\varepsilon}/\kappa)^{-1}}{ \sqrt{N}\sum_{i=1}^N u_i^2(\lambda_i+\sqrt{\varepsilon}/\kappa)^{-1} }  \\
&\leq&\displaystyle\left(\frac{\norm{\bar{\mathbf{r}}}_2}{2\kappa}+\frac{\lambda_{[1]}\left(\lambda_{[1]}-\lambda_{[N]}\right)}{\sqrt{N}\lambda_{[N]}}\right)\frac{1}{\lambda_{[N]}+\sqrt{\varepsilon}/\kappa},
\end{array}
\end{equation*}
where we repeatedly use the following relationships: $\norm{\mathbf{u}}_2^2=N$, $\norm{\widehat{\Lambda}^{-1}}_2\leq(\lambda_{[N]}+\sqrt{\varepsilon}/\kappa)^{-1}$, and $\frac{\lambda_{[1]}+\sqrt{\varepsilon}/\kappa}{\lambda_{[N]}+\sqrt{\varepsilon}/\kappa}\leq \frac{\lambda_{[1]}}{\lambda_{[N]}}$.

\section{Proof of Proposition 4}
When we increase $\delta$ by $\Delta$, the conditions that the optimal $s^\prime$ in ($\phi$, $\delta+\Delta$) is smaller than $s^\star$ in ($\phi$, $\delta$) are given by
\begin{equation*}
\left\{\begin{array}{l}
v_U(s^\star;\delta+\Delta, \phi)> v_U(l;\delta+\Delta, \phi), \exists\,l<s^\star,\\
v_U(k;\delta+\Delta, \phi)\geq v_U(s^\star;\delta+\Delta, \phi), \forall\,k\geq s^\star.
\end{array}\right.
\end{equation*}
Specifically, the first inequality guarantees that there exists a $l<s^\star$ such that $v_U(s^\star;\delta+\Delta, \phi)> v_U(l;\delta+\Delta, \phi)$, and the second inequality guarantees that any $s\geq s^\star$ is not the optimal solution. Similarly, the condition that $s^\prime$ is equal to $s^\star$ is given by
\begin{equation*}
v_U(l; \delta+\Delta, \phi)\geq v_U(s^\star;\delta+\Delta, \phi), \forall\,l,
\end{equation*}
and the conditions that $s^\prime$ is greater than $s^\star$ are given by
\begin{equation*}
 \left\{\begin{array}{l}
v_U(s^\star;\delta+\Delta, \phi)\geq v_U(l;\delta+\Delta, \phi), \exists\,l>s^\star,\\
v_U(k;\delta+\Delta, \phi)\geq v_U(s^\star;\delta+\Delta, \phi), \forall\,k\leq s^\star.
\end{array}\right.
\end{equation*}
Finally, we can derive $B_{\pm}(s^\star)$ by expanding above inequalities.

\section{Proof of Theorem \ref{th:localmin}}

We first prove result (i). Let $s\in\R$ satisfy the condition \eqref{cond-lift}.
Suppose that there exists $i\in\{1,\ldots,n\}$ such that $\bar\x_i\in (-t,0)\cup(0,t)$, then from Definition \ref{def-lifted} we have
\begin{equation}\label{prop-zero-eq1}
[\nabla h(\x)]_i+\frac{\phi_i}{t}+s=0\,\,\hbox{or}\,\,[\nabla h(\x)]_i-\frac{\phi_i}{t}+s=0,
\end{equation}
and there exists index $j\neq i$ such that $\bar\x_j\in(t,+\infty)$. Therefore, we know from Definition \ref{def-lifted}  that
$ [\nabla h(\bar\x)]_j+s=0$. This, together with \eqref{prop-zero-eq1}, implies that $\phi_i/t=|[\nabla h(\bar\x)]_i-[\nabla h(\bar\x)]_j|\leq 2L_h$, which contradicts to the condition $t<\{1/n,\phi_{\min}/2L_h\}$. As a consequence, result (i) holds.

For the result (ii), it holds naturally if $|\bar\x_i|\neq t$. When $|\bar\x_i|=t$, if $\bar{d}_i=0$, there exists $s\e\in\cN_{\cC}(\bar\x)$ such that
$[\nabla h(\bar\x)]_i+({\phi_i}/{t}){\rm sign}(\bar\x_i)+s=0$ according to Definition \ref{def-lifted}. The equality constraint $\e^T\x-1=0$ implies that there exists an index $j\neq i$ satisfying $\x_j>t$. Therefore, similar to the proof of result (i), it contradicts to the condition $t<\{1/n,\phi_{\min}/2L_h\}$. Then we can obtain result (ii).

Based on results (i) and (ii), results (iii) and (iv) can be derived
directly from Theorem 2.4 and Proposition 2.5 of \cite{Bian2020}.

\section{Proof of Proposition \ref{prop-KL}}

Note that $\varphi_t(s)=\frac{1}{t}[|s|-\max\{s-t,-s-t,0\}]$ can be equivalently written as $\varphi_t(s)=\min \{|s|/t,1\}$. Then there exist $2^n$  piecewise linear functions $P_i(\x)$ such that $\sum^{n}_{i=1} \phi_i\varphi_t(\x_i)=\sum^n_{i=1} \phi_i\min \{|\x_i|/t,1\}=\min_{1\leq i\leq2^n} P_i(\x)$. Additionally, the non-singularity of the matrix $W$ implies that $\cC=\{x: e^TW^{-1}Wx=1\}$. Therefore, the function $f$ can be reformulated into the following form:
$$
f(x)=l(Wx)+\min_{1\leq i\leq2^n} P_i(\x),
$$
where $l(y):=\frac{1}{2}\|y\|^2+\lambda\|y\|+\I_{\cC_y}(y)$ with $\cC_y=\{y: e^TW^{-1}y=1\}$. The fact $0\notin \cC_y$ implies that
the function $l$ is a proper closed convex function with an open domain, is strongly convex on any compact convex subset of $\dom~l$, and is twice continuously differentiable on $\dom~l$. The rest of the proof follows from Corollary 5.1 of \cite{Li2018}.

\section{Proof of Theorem \ref{th-convergence}}
From the convexity of functions $g_k$ and $q_t$, one has
$$
g_k(\x_k)\geq g_k(\x^{k+1})+\langle \delta_k,\x_k-\x^{k+1}\rangle\,\,\hbox{and}\,\,
q_t(\x^{k+1})\geq q_t(\x^{k})+\langle q^k,\x^{k+1}-\x^k\rangle.
$$
Then, we have
\begin{equation}\label{th-eq1}
\begin{array}{rl}
f(\x^{k+1})&=g_k(\x^{k+1})-q_t(\x^{k+1})+\langle q^k,\x^{k+1}-\x^k\rangle-\frac{\sigma_k}{2}\|\x^{k+1}-\x^k\|^2\\[2mm]
~ & \leq g_k(\x^k)+\langle \delta_k,\x^{k+1}-\x^k\rangle - q_t(\x^k)-\frac{\sigma_k}{2}\|\x^{k+1}-\x^k\|^2\\[2mm]
~ & \leq f(\x^k)+\|\delta_k\|\|\x^{k+1}-\x_k\|-\frac{\sigma_k}{2}\|\x^{k+1}-\x^k\|^2\\[2mm]
~& \leq f(\x^k)-\frac{\sigma_k}{4}\|\x^{k+1}-\x^k\|^2.
\end{array}
\end{equation}
This, together with the fact that $f(\x)\geq 0$, implies that
the sequence $\{f(\x^k)\}$ converges to a finite number.

Also from \eqref{th-eq1}, we have that
$$
0\leq \lim\limits_{k\to\infty} \frac{\sigma_k}{4}\|\x^{k+1}-\x^k\|^2\leq \lim\limits_{k\to\infty} [f(\x^k)-f(\x^{k+1})]\leq 0.
$$
{Since $f(\x)\to +\infty$ if $\|\x\|\to \infty$}, we know that the sequence $\{\x^k\}$ is bounded. Therefore, there exists a convergent subsequence $\{\x^k\}_{k\in\mathcal{K}}$ whose limit is $\x^{\infty}$.
Let $f_1(\x):=\frac{1}{2}\|W\x\|^2+\lambda \|W\x\|+ p_t(\x)+\I_{\mathcal{C}}(\x)$, then for any $k\in\mathcal{K}$,
$$
\begin{array}{l}
f_1(\x)-\langle q^k,x-\x^{k}\rangle +\frac{\sigma_k}{2}\|x-\x^{k}\|^2\\[2mm]
\geq f_1(\x^{k+1})-\langle q^k,\x^{k+1}-\x^{k}\rangle +\frac{\sigma_k}{2}\|\x^{k+1}-\x^{k}\|^2+\langle \delta_k,\x^{k+1}-\x^{k}\rangle\\[2mm]
\geq f_1(\x^{k+1})-\langle q^k,\x^{k+1}-\x^{k}\rangle +\frac{\sigma_k}{2}\|\x^{k+1}-\x^{k}\|^2- \|\delta_k\|\|\x^{k+1}-\x^{k}\|.
\end{array}
$$
{Suppose that $q^k\to q^{\infty}$ if $k \stackrel{\mathcal{K}}{\longrightarrow}+\infty$ (taking a subsequence of $\{\x^k\}_{k\in\mathcal{K}}$ is necessary).} By taking the limit $k \stackrel{\mathcal{K}}{\longrightarrow}+\infty$ of both sides of the above inequality, we have
$$
f_1(\x)-\langle q^{\infty},\x-\x^{\infty}\rangle +\frac{\sigma_{\infty}}{2}\|\x-x^{\infty}\|^2
\geq f_1(\x^{\infty}).
$$
This implies that $x_{\infty}$ is the optimal solution of the following convex optimization problem
$$
\min_{\x}~ f_1(\x)-\langle q^{\infty},\x\rangle +\frac{\sigma_{\infty}}{2}\|\x-\x^{\infty}\|^2.
$$
From its first-order optimality condition, we have
$0\in\partial f_1(\x^{\infty})-Q_{\infty}$. Since $q_t(\cdot)$ is a closed proper convex function, we can obtain from Theorem 24.4 in \cite{Rock1996} that $q^{\infty}\in\cQ(\x^{\infty})$.
Additionally, from Theorem \ref{th:localmin}  (i) and Definition \ref{def-lifted}, we know that $\x^{\infty}$ is a lifted stationary point of problem \eqref{uL2CP}. Consequently,  $\x^{\infty}$ is the local minimizer of RSMV from Theorem \ref{th:localmin} (ii). Furthermore, Proposition \ref{prop-KL} and Theorem 5 of \cite{Attouch2009} can guarantee the locally linearly convergence rate. The proof is completed.

\section{Semismooth Newton-CG method}\label{appendix-ssn}

In this part,  we present details of the semismooth Newton method for solving subproblem \eqref{alg-sub0} of the proximal dc algorithm. We first recall the explicit proximal mappings and their subdifferentials associated with the $\ell_2$ norm and $p_t$.
For any given $\lambda>0$,
\begin{equation}\label{def-prox2norm}
{\prox}_{\lambda\|\cdot\|}(\x)=
\left\{\begin{array}{ll}
\displaystyle\frac{\x}{\norm{\x}}(\norm{\x}-\lambda), & \hbox{\it if}\,\,\|\x\|>\lambda,\\
0, & \hbox{otherwise,}		
\end{array}
\right.
\end{equation}
and its generalized Jacobian is given by
\begin{equation}\label{def-prox2norm-Grad}
\partial {\prox}_{\lambda\|\cdot\|}(\x) =
\left\{
\begin{array}{ll}
\displaystyle \left\{I-\frac{\lambda}{\|\x\|} ( I - \frac{\x\x^T}{\|\x\|^2})\right\}, & \mbox{\it if } \|\x\| >  \lambda, \\[2mm]
\displaystyle \left\{t\frac{1}{\lambda^2}\x \x^T: 0\leq t\leq 1\right\}, & \mbox{\it if } \|\x\| = \lambda, \\[2mm]
\{0\}, & \mbox{\it if } \|\x\| <  \lambda.
\end{array}
\right.
\end{equation}
The proximal mapping associated with function $p_t$ at $\x\in\R^n$ can be characterized by
\begin{equation}\label{def-prop-L1}
[\prox_{p_t}(\x_i)]_i
=\left\{\begin{array}{ll}
\x_i+\phi_i/t, & \x_i<-\phi_i/t,\\[1mm]
0, & -\phi_i/t\leq \x_i\leq\phi_i/t,~~~~~i=1,\ldots,n,\\[1mm]
\x_i-\phi_i/t, & \x_i> \phi_i/t,
\end{array}
\right.
\end{equation}
and its generalized Jacobian is
\begin{equation}\label{def-prop-L1-Grad}
\partial\prox_{p_t}(\x)=\left\{\Theta\in\S^n:\Theta={\rm Diag}(\theta),\,\,\theta_i \in
\left\{\begin{array}{ll}
\{1\}, & \hbox{if $ \x_i< -\phi_i/t$ or $ \x_i> \phi_i/t$,}\\[1mm]
\{0\}, & \hbox{if $-\phi_i/t<  \x_i<\phi_i/t$,}~~~~~~~~~~i=1,\ldots,n\\[1mm]
[0,1], & \hbox{if $ \x_i=-\phi_i/t$ or $ \x_i= \phi_i/t$,}
\end{array}
\right.
 \right\}.
\end{equation}
The $k$-th subproblem \eqref{alg-sub0} can be equivalently written as
 \begin{equation} \label{alg-sub-1}
 \begin{array}{cl}
\min\limits_{\x,u} &\frac{1}{2}\|u\|^2+\lambda \|u\|+ p_t(\x)-\langle q^k,\x-\x^k\rangle+\frac{\sigma_k}{2}\|\x-x^k\|^2\\
{\rm s.t.} & W\x-u=0,\\
~& \e^T\x-1=0.
\end{array}
\end{equation}
The Lagrangian function associated with problem \eqref{alg-sub-1} is given by
$$
\bL_k(\x,u,\y,v) = \frac{1}{2}\|u\|^2+\lambda \|u\|+ p_t(\x)-\langle Q_k,\x-\x^k\rangle+\langle W\x-u,\y\rangle+\langle \e^T\x-1,v\rangle +\frac{\sigma_k}{2}\|\x-\x^k\|^2.
$$
By strong duality theorem (see,e.g., Theorem 36.3 of \cite{Rock1996}), we have
\begin{equation}\label{minmax}
\begin{array}{l}
\min\limits_{\x,u}\max\limits_{\y,v}~\bL_k(\x,u,\y,v) \\
= \max\limits_{\y,v} \min\limits_{\x,u}~\bL_k(\x,u,\y,v) \\
= \max\limits_{\y,v} ~\left\{\min\limits_u\{\lambda \|u\|-\langle u,\y\rangle+\frac{1}{2}\|u\|^2\}+\min\limits_{\x} \{p_t(\x)+\langle W^T\y+\e v-q^k,\x\rangle+\frac{\sigma_k}{2}\|\x-\x^k\|^2\}-v\right\}\\[2mm]
=\max\limits_{\y,v}  \Big\{\mathcal{M}_{\lambda\|\cdot\|}(\y)-\frac{1}{2}\|\y\|^2-v\\
~~~~~~~~~+\sigma_k\mathcal{M}_{p_t/\sigma_k}\big(\x^k-(W^T\y+\e v-q^k)/\sigma_k\big)
-\frac{\sigma_k}{2}\|\x^k-(W^T\y+\e v-q^k)/\sigma_k\|^2+\frac{\sigma_k}{2}\|\x^k\|^2\Big\}.
\end{array}
\end{equation}
Then we can obtain the objective function of the dual problem associated with problem \eqref{uL2CP}:
$$
\max\limits_{\y,v}~-h_k(\y,v),
$$
where $h_k$ is given by \eqref{def-h}.

\section{Proof of Theorem \ref{prop-SN}}\label{appendix-th3}
To establish the convergence of the proposed semismooth Newton method, we first present Lemmas 1 and 2. In particular, Lemma 1 can be obtained directly from Lemma 2.1 in \cite{Zhang2020} and Theorem 3.6 in \cite{Li2018}, and hence its proof is omitted.
\begin{lemma}\label{lemma1}
Let the multifunction $\mathcal{G}_k:\R^{n+1}\rightrightarrows\mathbb{S}^{n+1}$ be defined as \eqref{def-G}.
Then, one has
\begin{itemize}[leftmargin=6mm]
\item[(a)] the multifunction $\mathcal{G}_k$ is  nonempty compact valued upper-semicontinuous;
\item[(b)] any element in $\mathcal{G}_k(\y,v)$ is positive semidefinite;
\item[(c)] $\nabla h_k$ is strongly semismooth on $\R^n$ with respect to $\mathcal{G}_k$, i.e., $\nabla h_k$ is directionally differentiable at $(\y;v)$ and for any $G\in\mathcal{G}_k(\y+d_y,v+d_v)$ with $d:=(d_y;d_v)\to 0$, it holds that
    $$
    \nabla h_k(\y+d_y,v+d_v)-\nabla h_k(\y,v)-Gd=\mathcal{O}(\|d\|^2).
    $$
\end{itemize}
\end{lemma}

\begin{lemma}\label{lemma2}
Suppose that $\hat{\y},\hat{v}$ satisfies $\nabla h_k(\y,v)=0$, then all the elements in $\mathcal{G}_{k}(\hat{\y},\hat{v})$ are symmetric and  positive definite.
\end{lemma}
{\it Proof. } It follows from  \eqref{def-prop-L1} and \eqref{minmax} that the optimal solution $(\hat{\x},\hat{u})$ to problem \eqref{k-dual} satisfies $\hat{x}=\prox_{p_t/\sigma_k}(\tilde{x}^k(\hat{\y},\hat{v}))$ and $\hat{u}=W\hat{\x}= \prox_{\lambda\|\cdot\|}(\hat{\y})$. Therefore, the positive definiteness of matrix $W$ implies that $\hat{u}\neq 0$. This, together with \eqref{def-prox2norm}, shows $\|\hat{\y}\|>\lambda$.
Consequently, we can obtain from \eqref{def-prox2norm-Grad} that $U\in\partial\prox_{\lambda\|\cdot\|}(\hat{\y})$ is positive definite.
 Furthermore, $\e^T\prox_{p_t}(\tilde{x}^k(\hat{y},\hat{v}))=1$ and \eqref{def-prop-L1} imply that there exists $i\in\{1,\ldots,n\}$ such that $\tilde{x}^k_i(\hat{y},\hat{v})<-\frac{1}{t}\phi_i$ or $\tilde{x}^k_i(\hat{y},\hat{v})>\frac{1}{t}\phi_i$. Then, for any $V\in\partial\prox_{p_t}(\tilde{x}^k(\hat{\y},\hat{v}))$,
we have  $V\neq 0$ and $\e^TV\e>0$.
From the positiveness of $U$ and $W$, we know that there exists a positive scalar $\alpha$ such that
$\sigma_kU-\alpha WW^T\succ0$ and
$$
W^T(\sigma_kU+ WVW^T)^{-1}W\prec W^T(W(\alpha I+V)W^T)^{-1}W=(\alpha I+V)^{-1}.
$$
Then, the Schur complement of $\sigma^{-1}_k\e^TV\e$ in $G$ satisfies
$$
\sigma^{-1}_k\e^TV\e- \sigma^{-2}_k\e^TVW^T(U+\sigma^{-1}_k WVW^T)^{-1}WV\e>\sigma^{-1}_k \left[\e^TV\e- \sigma^{-2}_k\e^TV(\alpha I+V)^{-1}V\e \right]\geq 0.
$$
This completes the proof of Lemma \ref{lemma2}.

With results of Lemmas \ref{lemma1} and \ref{lemma2}, the proof of Theorem \ref{prop-SN} follows directly from Theorems 3.4 and 3.5 of \cite{Zhao2010}.

\end{appendix}


	
\end{document}